\documentstyle[12pt]{article}

\newcommand{\rap}{ \mbox{$\frac{\epsilon'}{\epsilon}$} }
\newcommand{\rapp}{ \mbox{$\frac{\epsilon'}{\epsilon},$} }
\newcommand{\DAFNE}{DA$\Phi$NE}
\newcommand{\dis}{\displaystyle}
\begin{document}
\begin{titlepage}

\begin{flushright}
Rome-preprint-1071\\
CERN-TH-7504\\
Nov. 1994\\
\end{flushright}
\vspace{2 cm}

\begin{center}
{\Large  {\bf CP and CPT  measurements at \DAFNE$^{*}$ } }\\
      \vspace{2 cm}
    {\bf G. D'Ambrosio}\\
      \vspace{.3 cm}
        TH-Division, CERN, Geneva, Switzerland$^{**}$\\
     \vspace{2 cm}
    {\bf G. Isidori and A. Pugliese} \\
      \vspace{.3 cm}
        Dipartimento di Fisica, Universit\'a di Roma "La Sapienza" \\
        and \\
        INFN - Sezione di Roma, Rome, Italy
    \vspace{2 cm}
\end{center}

\begin{abstract}
Starting from the time evolution of the C-odd $\bar{K}^0 K^0$  system, we
analyze the asymmetries measurable at \DAFNE~ and their implications on
CP violation and on the possible tests of T and CPT symmetries. In
particular the ratio \rap~ can be measured with high precision (up to about
$10^{-4}$ for the real part). The CP-, T- and CPT-violating parameters
can be explored in $K_S$ semileptonic
decays with an accuracy of the order of $10^{-3}$. The possibility to detect
 $K_S \to 3 \pi$ and $K_L \to \pi \pi \gamma$ decays is also discussed.
\end{abstract}

\vfill
\noindent {\small $^*$ Work supported in part by HCM, EEC--Contract No.
            CHRX--CT920026 (EURODA$\Phi$NE) }

\noindent {\small $^{**}$ On leave of absence from INFN - Sezione di
    Napoli, Naples, Italy }
\thispagestyle{empty}
\end{titlepage}

\newpage

\section{Introduction}

The $\bar{K}^0$ $K^0$ state produced in the decay of the $\phi$
resonance is odd under charge conjugation and is therefore an antisymmetric
$K_{L}$ $K_{S }$ state.  This characteristic makes a $\phi$ factory very
suitable to study  CP violation and to test CPT symmetry in $K$ meson
decays \cite{vign,ohs}.

For a long time it has been stressed that
the presence in the same detector of $K_{L}$ and $K_{S}$ beams,
produced without regeneration and thus with the
relative fluxes perfectly known, will allow a very clean determination of the
ratio $\rap$ \cite{franzini}.
A non-zero value for $\rap$ is an unambiguous signal of
the existence of direct CP violation, which is naturally expected in the
Standard Model. The present experimental situation is:
\begin{eqnarray}
\Re\left(\rap\right) = \left( 2.3 \pm 0.7 \right) \times 10^{-3}
	& &
\Im\left(\rap\right) = \left( -1.2 \pm 17 \right) \times 10^{-3}
		\hspace{1.5 cm}{\rm NA}31\cite{burk} \nonumber \\
\Re\left(\rap\right) = \left( 0.74 \pm 0.60 \right) \times 10^{-3}
	& &
\Im\left(\rap\right) = \left( +4.7 \pm 3.5 \right) \times 10^{-3},
		\hspace{1.5 cm}{\rm E}731\cite{pat} \nonumber
\end{eqnarray}
still consistent with $\epsilon' = 0$.
The theoretical calculation of $\rap$ in the Standard Model
is strongly affected by QCD corrections and, for large values
of the top mass, large cancellations are expected \cite{franco}.
The present estimate is $\Re(\rap)=(2.8\pm 2.4)\times10^{-4}$
\cite{franco2}, thus a fundamental goal of \DAFNE\
is certainly to reach the sensitivity of $10^{-4}$ in the measurement
of this ratio. Independent information about direct CP violation
could be obtained also by charge asymmetries in
$K^{\pm} \to 3\pi$ \cite{paver}
and $K^{\pm} \to 2\pi\gamma$ \cite{dibP} decays.
\par

Beyond the study of direct CP violation,
the presence of  a pure $ K_{S}$ beam will allow
the observation at \DAFNE\ of some suppressed $ K_{ S }$ decays,
such as the semileptonic and the three-pion ones.
The theoretical predictions for semileptonic
decays are not strongly affected by QCD corrections and, as
we will discuss later, the measurement of the semileptonic rates and
charge asymmetries in $K_{S,L}$ decays can give many interesting
tests of CPT and of the $\Delta S =  \Delta Q$ rule. Moreover, due to the
coherence of the initial state, T and CPT symmetries can be
directly tested in events with two leptons in the final state
\cite{bucha,fuku}.

Some time ago it was pointed out that the radiative decay
$\phi \to \gamma \, f_0  \to \gamma \, (K^0 \bar K^0)_{C=+} $
could have a non-negligible branching ratio \cite{cevena} and therefore
a dangerous background, namely a $K^0 \bar K^0$ component even
under charge conjugation, could be present.
New determinations of the $\phi  \to \gamma \, (K^0 \bar K^0)_{C=+} $
 branching ratio  \cite{cevenb} turn
out to be much smaller, then, as we will show, the inclusion of
 the C-even background does not sensibly affect
 the measurements of  \rap and $K_S$ suppressed decays.

Recently it as been suggested that quantum mechanics
violations may be generated by non-local theories
at the Planck scale \cite{QMV}. As a consequence
CP- and CPT-violating effects could be induced \cite{ellis}.
As we will discuss, the coherence of the $\phi$-factory
initial state will help in disentangling
these effects, and quite stringent bounds could be obtained for
the quantum mechanics violating parameters \cite{huet}.
It is worth while to note that the quantum mechanics violation induces
a loss of the initial state coherence which can somehow
simulate the effect of a C-even background.

\par
The plan of the paper is the following: in section 2 we recall the
time evolution of the initial state. In section 3 we report its implications
 on the determination of real and imaginary parts of \rap.
 In section 4 the
semileptonic decays, with possible direct tests of
T and CPT symmetry, are discussed. Sections 5 and 6 are devoted, respectively,
to $K_S\to 3 \pi$ and $K_L\to  \pi \pi \gamma$  decays.
In section 7 we
study the effect of the C-even background. Finally, in
 section 8 we discuss the implications of
possible quantum mechanics violations.

\section{Time evolution  }

The antisymmetric $\bar{K}^0 K^0$ state
produced in the $\phi$ decay can be written as
\begin{equation}
\phi \rightarrow {1\over\sqrt{2}}\left[ \bar{K}^{0(q)}K^{0(-q)} -K^{0(q)}
\bar{K}^{0(-q)} \right] =
 \frac{h}{\sqrt{2}}\left[ K_{ S }^{(q)}K_{ L }^{(-q)} -
K_{L}^{(q)}K_{ S }^{(-q)} \right],
\label{phid}
\end{equation}
where $q$ and $-q$ indicate the spatial momenta of the two-kaons and
the normalization factor $h$ is
\begin{equation}
h= \frac{1+ |\epsilon|^2}{1- \epsilon^2}\simeq 1.
\end{equation}
The decay amplitude of the two-kaon system into the final state
$|a^{(q)}(t_1),b^{(-q)}(t_2)\rangle$ is given by:
\begin{eqnarray}
A\left(a^{(q)}(t_1),b^{(-q)}(t_2)\right)&=&{h\over {\sqrt2}}\Big[ A(K_S\to
a) e^{-i\lambda_S t_1} A(K_L\to b)e^{-i\lambda_L t_2} \nonumber \\
& &\qquad -A(K_L\to a)e^{-i\lambda_L t_1}A(K_S\to b)e^{-i\lambda_S t_2} \Big],
\label{ampd}
\end{eqnarray}
where $\lambda_{S(L)}=m_{S(L)}-i\Gamma_{S(L)}/2$.
As usual we define also:
\begin{equation}
\Gamma={\Gamma_S+\Gamma_L\over 2},\qquad\qquad \Delta\Gamma=\Gamma_S-\Gamma_L
\qquad {\rm and}\qquad \Delta m=m_L-m_S.
\end{equation}

If $|a\rangle\not=|b\rangle$ the two states
$|a^{(q)}(t_1),b^{(-q)}(t_2)\rangle$
and $|a^{(-q)}(t_1),b^{(q)}(t_2)\rangle$ are physically different for any
value of $t_1$ and $t_2$, therefore the double differential rate is:
\begin{eqnarray}
\Gamma(a(t_1),b(t_2)) &=& |h|^2 \int \Big\lbrace |A_{S}^{a}|^2
|A_{L}^{b}|^2
e^{-(\Gamma_St_1+\Gamma_Lt_2)}    +  |A_{L}^{a}|^2 |A_{S}^{b}|^2
e^{-(\Gamma_Lt_1+\Gamma_St_2)}  \nonumber \\
& &\qquad -2\Re\left[ A_S^a A_L^{a*} A_L^b A_S^{b*}
e^{-\Gamma (t_1+t_2)+i\Delta m(t_1-t_2)}\right] \Big\rbrace d\phi_a
d\phi_b, \label{gamab}
\end{eqnarray}
where $\phi_a$ and $\phi_b$ are the phase spaces of the final states.
Integrating   eq. (\ref{gamab}) on $t_1$ and $t_2$ one obtains
the probability for the decay into the $|a,b\rangle$ state with both the
decay vertices inside the detector:
\begin{eqnarray}
 P(a,b) & = & {|h|^2 \over \Gamma_S \Gamma_L} \left[ \left(
 \Gamma_{S}^{a}\Gamma_{L}^{b}   +  \Gamma_{L}^{a}\Gamma_{S}^{b} \right) S_L
 -2{\Gamma_S\Gamma_L \over \Gamma^2 +\Delta m^2}
\int \Re\left( A_S^a A_L^{a*} A_L^b A_S^{b*}\right) d\phi_a d\phi_b
 \right]
\end{eqnarray}
where $S_L=(1-e^{-D/d_L})$ is the $ K_{L}$ acceptance of
the detector:  the KLOE project quotes for the fiducial length
  $ D\simeq 120$ cm \cite{franzini}
 ($d_L=340$ cm is the $K_L$ mean decay path), thus $S_L\simeq 0.3$.\par
For $|a\rangle= |b\rangle$ the interchange of $q \leftrightarrow -q$ is
equivalent
to  $t_1 \leftrightarrow t_2$, thus:
\begin{equation}
 P(a,a)  = |h|^2 { \Gamma_{S}^{a}\Gamma_{L}^{a}\over \Gamma_S \Gamma_L}
\left[ S_L  -{\Gamma_S\Gamma_L \over \Gamma^2 +\Delta m^2}
\right].
\end{equation}
As   will be discussed in the following, the choice of appropriate time
integration
intervals  supplies a powerful tagging of $K_L$ or $K_S$ decays.
\par
Finally we define also the so-called ``time difference distribution'':
\begin{eqnarray}
 &&I(a,b;t) = \int dt_1 dt_2 |A (a(t_1),b(t_2))|^2 \delta(t_1-t_2-t)
\nonumber \\
  &&= {|h|^2 e^{-\Gamma|t|} \over 2\Gamma} \Big\lbrace
|A_{S}^{a}|^2|A_{L}^{b}|^2 e^{-{\Delta\Gamma\over 2}t}
+|A_{L}^{a}|^2|A_{S}^{b}|^2 e^{+{\Delta\Gamma\over 2}t}
  -2\Re\left[ A_S^a A_L^{a*} A_L^b A_S^{b*}
e^{+i\Delta m t}\right] \Big\rbrace.
\label{intt}
\end{eqnarray}

\section{Real and imaginary parts of  \rap }
\par
As extensively  discussed, for example in Refs. \cite{franzini,duni,patera},
the study of the time difference distribution, for
$\pi^+ \pi^-$, $\pi^0\pi^0$  final states,
leads to the determination of both
$\Re\left(\rap\right)$ and $\Im\left(\rap\right)$.\par
Introducing as usual the amplitudes
\begin{equation}
 \eta^{+-}  = \frac{ A \left( K_{L}\rightarrow \pi^{+} \pi^{-}\right) }
                       { A\left( K_{S}\rightarrow \pi^{+} \pi^{-}\right) }
             =  \epsilon + \epsilon'  \qquad {\rm and} \qquad
 \eta^{00}  =  \frac{ A \left( K_{L}\rightarrow \pi^{0} \pi^{0}\right)}
                       { A\left( K_{S}\rightarrow \pi^{0} \pi^{0}\right)}
             = \epsilon - 2\epsilon',
\end{equation}
eq. (\ref{intt}), integrated over the pion phase space, gives:
\begin{eqnarray}
F(t)  &= & \int d\phi_{+-}\, d\phi_{00}\, I(\pi^+ \pi^-,\pi^0 \pi^0;t)
\nonumber \\ & =&
                   \frac{ \Gamma_{S}^{+-}\Gamma_{S}^{00}}
{2 \Gamma}e^{-\Gamma|t|}
              \left [\mid\eta^{+-}\mid^{2}e^{+{\Delta\Gamma\over 2}t}
+\mid\eta^{00}\mid^{2} e^{-{\Delta\Gamma\over 2}t}
            -2\Re\left ( \eta^{+-}(\eta^{00})^{*}
e^{-i\Delta m t}\right)\right ].
\label{epsp}
\end{eqnarray}
\par

If $ \epsilon' \neq 0$ there is an asymmetry between the events
 with positive and negative values of $ t$:
\begin{equation}
    A\left(t\right)  =
    \frac{F\left(|t| \right) - F\left( - |t| \right)}
         {F\left(|t| \right) + F\left(- |t| \right)} =
        A_{R}\left( t\right) \times \Re\left(\rap\right) -
        A_{I}\left( t\right) \times \Im\left(\rap\right),
\label{aepsp}\end{equation}
neglecting in eq. (\ref{aepsp}) terms proportional to
$ \left(\frac{\epsilon'}{\epsilon}\right)^{2}$,
the $ A_{R}\left( t\right)$
and $ A_{I}\left( t\right)$ coefficients, shown in Fig.1, are given by:
\begin{equation}
\begin{array}{rcl}
    A_{R}\left( t\right)
            & = &
    3\dis\frac{
            e^{-|t|\Gamma_L} -
            e^{-|t| \Gamma_S}
          } {
            e^{ -|t| \Gamma_L } +
            e^{ -|t|\Gamma_S } -
            2 \cos\left( \Delta m|t| \right)
            e^{ -\Gamma |t| }
           }  \\  & & \\
    A_{I}\left( t\right)
            & = &
    3\dis\frac{
            2\sin\left( \Delta m |t| \right)
            e^{-\Gamma|t|}
           } {
            e^{ -|t| \Gamma_L } +
            e^{ -|t|\Gamma_S } -
            2 \cos\left( \Delta m|t| \right)
            e^{ -\Gamma |t| }
           }.
\end{array}
\label{arai}
\end{equation}
\par
It can be seen that $ A_{ R }(t)$ becomes nearly
independent of $ t$, and equal to 3, for $ t>>\tau_{ S }$; on
the other hand $ A_{ I }(t)$ is strongly dependent on $ t$ and
vanishes for $ t>>\tau_{ S }$.
 \begin{figure}[t]  
     \begin{center}
        \setlength{\unitlength}{1truecm}
        \begin{picture}(8.0,8.0)
\put(-4.0,-6.0){\special{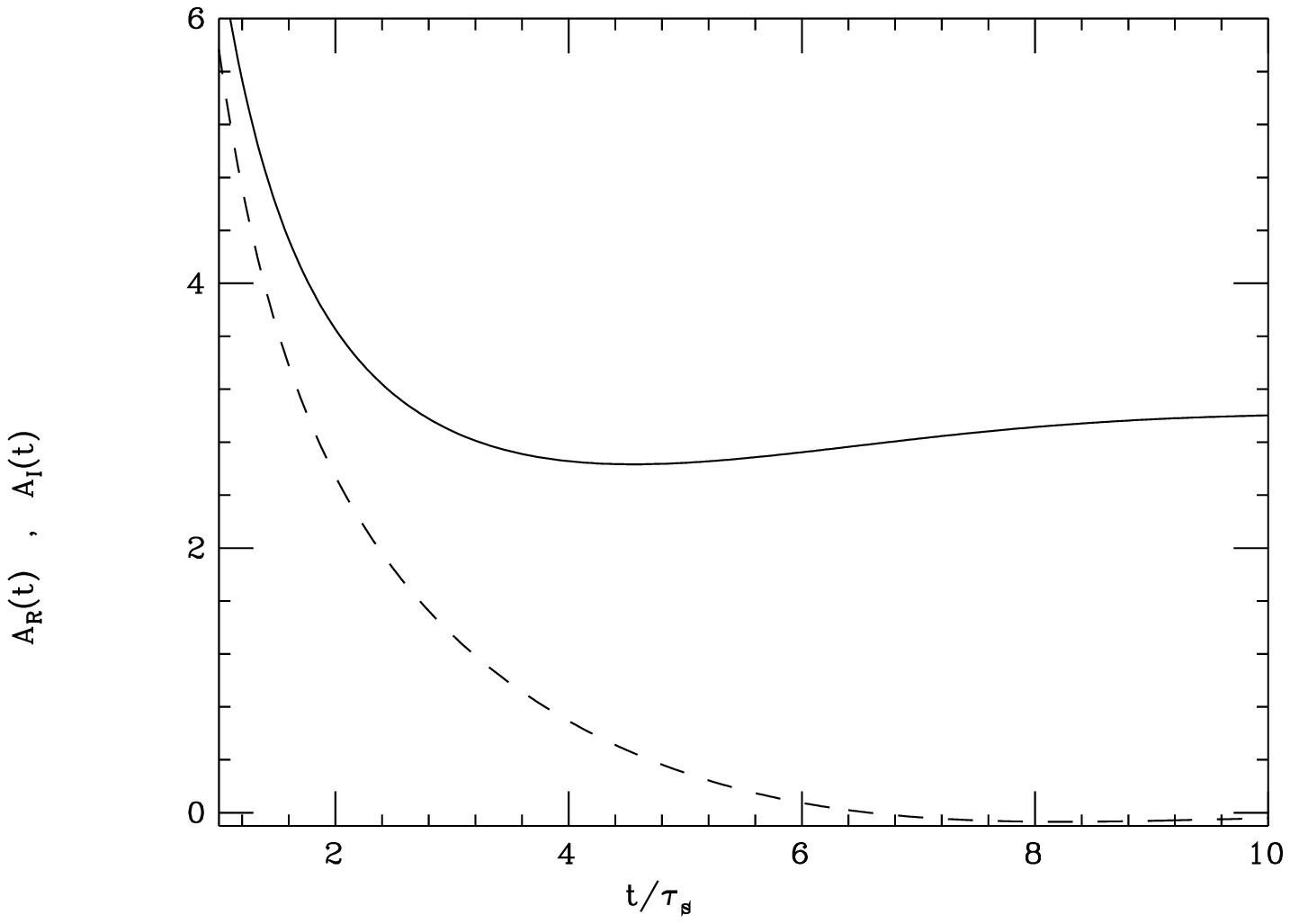}}
        \end{picture}
    \end{center}
    \caption{Coefficients of $\Re\left(\rap\right)$ (full line)
and $\Im\left(\rap\right)$ (dashed line) defined in eq.(11).}
\end{figure}
Therefore a measurement of the asymptotic value
of $ A(t)$ or of the value of the integrated asymmetry
\begin {equation}
A = \frac{F(t>0) - F(t<0)}{F(t>0) + F(t<0)}\simeq
3\Re\left(\frac{\epsilon'}{\epsilon}\right)
\end{equation}
allows a clean determination of
$\Re\left(\frac{\epsilon'}{\epsilon}\right)$. The statistical error on $A$
is
given by:
\begin{equation}
\sigma_{A} = \sqrt{ \frac{\left( 1+A \right)\times \left( 1-A \right) }
                        { N } },
\end{equation}
where $N$ is the number of $\phi \rightarrow \pi^{+}\pi^{-},\pi^{0}\pi^{0}$
events.
At the reference \DAFNE~ luminosity the  statistical error on
$\Re\left(\rap\right)$ is then:
\begin{equation}
 \sigma_{\Re\left(\rap\right)} \simeq
        \frac{\sigma_{A}}{3} \simeq \frac{1}{3\sqrt{N}} \simeq 1.7 \times
        10^{-4}.
\end{equation}

The integrated asymmetry $A$ allows a precise determination of
$\Re \left( \rap \right)$ but gives no information on the imaginary part of
\rap.
To overcome this problem a further method can be exploited to measure both
$\Re\left(\rap\right)$ and $\Im\left(\rap\right)$ from the
$K_{L}K_{S}\rightarrow \pi^{0}\pi^{0},\pi^{+}\pi^{-}$ decay time difference:
the experimental distribution $F\left(d\right)$ \footnote{ The decay times
are measured through the decay paths, and the time difference $t$ is given
by  $t = (d_c - d_n) \tau_S /d_S  = d \, \tau_S /d_S $ ,
 where $d_c$ ($d_n$) is the
 decay path into charged (neutral) pions and $d_S \simeq 0.6$ cm
is the mean decay path of the $K_S$. }
 can be fitted by the
theoretical distribution of eq. (\ref{epsp}), and $\Re\left(\rap\right)$ and
$\Im\left(\rap\right)$ can be used as free parameters of the fit.
\par
It must be stressed that this procedure is very sensitive to the
experimental resolution on the measurement of $d$.
 The information
contained in the shape of the
$F\left(d\right)$ distribution can be easily washed out, in particular
in the region of interest for the determination of $\Im\left(\rap\right)$,
where $d \simeq d_{S}$. In fact only in this range of  $d$
values can $A_{I}\left( d \right)$ be different from zero and the strongly
varying
behaviour of $A_{I}\left( d \right)$   smeared out by a bad vertex
reconstruction.
This effect is shown in Fig. 2, where the theoretical distribution is compared
with a simulated experimental distribution with a Gaussian error on the $d$
measurement equal to 5 mm.
\par
\begin{figure}
        \setlength{\unitlength}{1truecm}
        \begin{picture}(14.0,12.0)
        \special{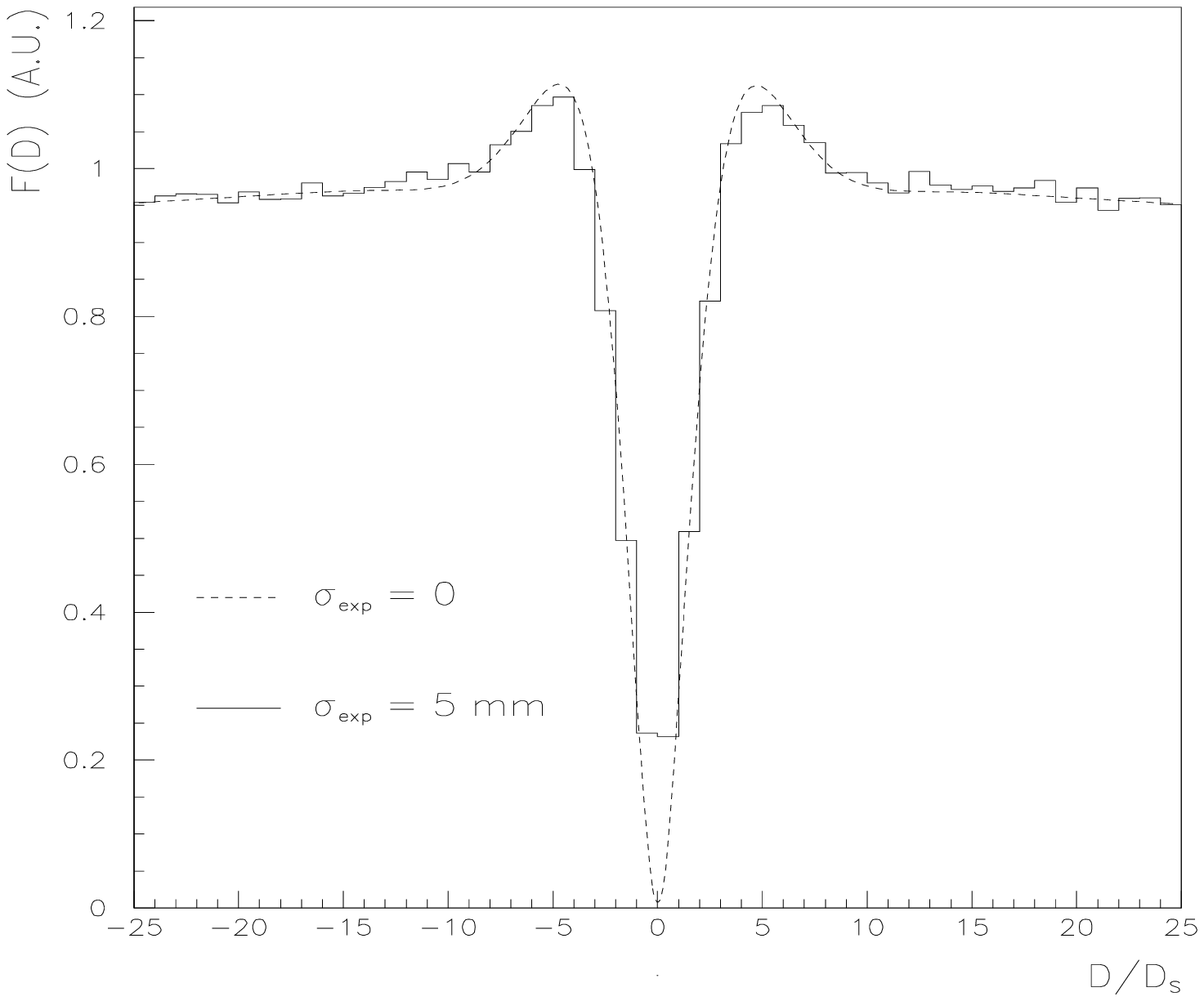}
        \end{picture}
   \caption{ Comparison between the theoretical F(d) distribution
for $\Re\left(\rap\right) = 2.8\times 10^{-4}$  and that
            obtained with an experimental vertex resolution $\sigma =
5{\rm mm}$.}
\end{figure}

The effects of the finite experimental resolution have been
discussed, for example in \cite{patera}, to which we refer.
The results of the quoted analysis are that the determination of
$\Re\left(\rap\right)$ is practically unaffected by the experimental
 resolution,
while the statistical error on $\Im\left(\rap\right)$ increases by more
than a factor $2$.
This analysis estimates that the accuracy achievable
for a realistic detector is:
\begin{equation}
    \sigma_{ \Re\left(\rap\right) } = 1.8 \times 10^{-4}~~~~~~~  ;~~~~~~~
    \sigma_{ \Im\left(\rap\right) } = 3.4 \times 10^{-3}.
\end{equation}
These numbers have to be compared with the present experimental situation
shown in the introduction.

\section{Semileptonic Decays}
\subsection{Theoretical introduction}

We will discuss the semileptonic decays of neutral kaons in a very
general framework, without assuming the $ \Delta S = \Delta Q$ rule
and the CPT symmetry. \par
The $ \Delta S = \Delta Q$ rule is well supported by experimental
data and is naturally accounted for by the Standard Model, where the $ \Delta
 S = - \Delta Q$ transitions are possible only with two  effective
weak vertices. Explicit
calculations give a suppression factor of about $10^{-6}$--$10^{-7}$
\cite{guberina}.
Furthermore in any quark model, $ \Delta S = - \Delta Q $
transitions can be induced only by operators with dimension
higher than 6 and therefore are suppressed \cite{maiani}. \par
Although it is very unlikely to have a theory with a large violation of
the $ \Delta S = \Delta Q$ rule, this does not conflict with any general
principle. On the contrary CPT symmetry
must hold in any Lorentz-invariant local field theory.
The problem of possible sources of CPT violation has recently received
much attention. Attempts to include also gravitation in the unification
of fundamental interactions lead to non-local theories, like superstrings,
which suggest possible CPT violation above the Planck mass, which turns out
to be the natural suppression scale \cite{peccei}. \par


We neglect for the moment quantum mechanics violating effects \cite{QMV,ellis},
which will be discussed
later, introducing CPT violation through an ``ad hoc"
parametrization of the decay amplitudes and
the mass matrix elements.
\par
Following the notations of Ref. \cite{maiani} we define:
\begin{eqnarray}
   A(K^0 \to l^+\nu \pi^-) &=& a + b \nonumber \\
   A(K^0 \to l^-\nu \pi^+) &=& c +d  \nonumber \\
  A(\bar{K}^0 \to l^-\nu \pi^+) &=& a^* -b^* \nonumber \\
  A(\bar{K}^0 \to l^+\nu \pi^-) &=& c^* -d^*
\label{ampkl}
\end{eqnarray}
CPT implies $ b=d=0$, CP implies $\Im(a)=\Im(c)=\Re(b)=\Re(d)=0$,
T requires real amplitudes and $ \Delta S = \Delta Q$ implies $c=d=0$.

Writing the mass matrix for the $K^0 \bar{K}^0$ system in the form:
\begin{equation} \left(\begin{array}{cc}
M_{11} - i\Gamma_{11}/2 &
M_{12} - i\Gamma_{12}/2 \\ & \\
M_{21} - i\Gamma_{21}/2  &
M_{22} - i\Gamma_{22}/2
\end{array}\right), \label{massm} \end{equation}
the eigenstates are given by:
\begin{equation}
\begin{array}{rcl}
    K_S & = &
        \dis\frac{1}{\sqrt{2\left( 1 + |\epsilon_S|^2\right)}}
                  \left[ \left(1 + \epsilon_S\right)K^0 +
                  \left( 1 - \epsilon_S \right) \bar{K}^0\right]   \\
    K_L & = &
        \dis\frac{1}{\sqrt{2\left( 1 + |\epsilon_L|^2\right)}}
                    \left[ \left(1 + \epsilon_L\right) K^0 -
                    \left( 1 - \epsilon_L \right) \bar{K}^0 \right],

\end{array}
\label{kskl}
\end{equation}
where the $\epsilon_i$ parameters are:
\begin{eqnarray}
\epsilon_S & = & \frac{
    -i \Im\left( M_{12}\right) -
            \frac{1}{2} \Im\left(\Gamma_{12}\right) -
    \frac{1}{2} \left[ M_{11} - M_{22} -\frac{i}{2}
    \left( \Gamma_{11} -\Gamma_{22}\right) \right]
                    }{
    m_L - m_S +i(\Gamma_S - \Gamma_L)/2
                    } = \epsilon_M +\Delta \nonumber  \\
 & & \label{esel} \\
\epsilon_L & = & \frac{
    -i \Im\left( M_{12}\right) -
            \frac{1}{2} \Im\left(\Gamma_{12}\right) +
    \frac{1}{2} \left[ M_{11} - M_{22} -\frac{i}{2}
    \left( \Gamma_{11} -\Gamma_{22}\right) \right]
                    }{
    m_L - m_S +i(\Gamma_S - \Gamma_L)/2
                    } = \epsilon_M -\Delta. \nonumber
\end{eqnarray}
Then the masses and widths are:
\begin{equation}
m_{S(L)} = \frac{  M_{11} +M_{22}  }{2} \pm \Re(M_{12}),\qquad\qquad
\Gamma_{S(L)} = \frac{  \Gamma_{11} +\Gamma_{22}  }{2} \pm \Re(\Gamma_{12}).
\label{msml}
\end{equation}
CPT symmetry would require $M_{11} = M_{22}$ and
$\Gamma_{11} = \Gamma_{22}$, recovering the relation
$\epsilon_S = \epsilon_L = \epsilon_M$.
\par
Using eqs. (\ref{ampkl}) and (\ref{kskl})
the semileptonic partial rates are given by:
\begin{eqnarray}
\Gamma_S ^{l^\pm} &=& {|a|^2 \over 2}\left[1 \pm 2\Re(\epsilon_S) \pm 2 \Re
\left({b \over a}\right) + 2 \Re\left({c^* \over a}\right) \mp
2\Re\left({d^* \over a}\right)\right] \nonumber \\
& & \label{gsgl} \\
\Gamma_L ^{l^\pm} &=& {|a|^2 \over 2} \left[1 \pm 2\Re(\epsilon_L) \pm 2 \Re
\left({b \over a}\right) - 2 \Re\left({c^* \over a}\right) \pm
2\Re\left({d^* \over a}\right)\right]; \nonumber
\end{eqnarray}
thus the charge asymmetries for $K_S$ and $K_L$ are:
\begin{eqnarray}
\delta_S &=& \frac{  \Gamma_S ^{l^+} - \Gamma_S ^{l^-}}{ \Gamma_S ^{l^+}+
 \Gamma_S ^{l^-}} =  2\Re(\epsilon_S) + 2 \Re\left({b
\over a}\right) -2\Re\left({d^* \over a}\right) \nonumber \\
& & \label{deltasl} \\
\delta_L &=& \frac{  \Gamma_L^{l^+} - \Gamma_L ^{l^-}}{ \Gamma_L ^{l^+}+
 \Gamma_L ^{l^-}} =  2\Re(\epsilon_L) + 2 \Re\left({b
\over a}\right) +2\Re\left({d^* \over a}\right). \nonumber
\end{eqnarray}
A non-vanishing value of the
difference $\delta_S -\delta_L$ would be an evidence of
CPT violation, either in the
mass matrix or in the ${\Delta S=-\Delta Q}$ amplitudes ($ \Delta$ and
$ d^*/a $ cannot be disentangled by semileptonic decays alone).
The sum $\delta_S +\delta_L$ has CPT-conserving ($\Re(\epsilon_M)$) and
CPT-violating ($\Re (a/b)$) contributions that cannot be disentangled.
\par
The ratio of  $K_S$ and $K_L$ semileptonic widths
\begin{equation}
\eta={\Gamma_{S}^{l} \over \Gamma_{L} ^{l}} =
1+4 \Re\left({c^*\over a}\right),
\label{eta}
\end{equation}
where $\Gamma_{S(L)} ^{l}=\Gamma_{S(L)} ^{l^+}+\Gamma_{S(L)} ^{l^-}$,
allows us to determine the
CPT-conserving part of the amplitudes with $\Delta S = -\Delta Q$.
\par
All imaginary parts disappear from the rates and only the time evolution
can potentially give some information on them.
\par

\subsection{Determination\ \ of\ \ semileptonic\ \ branching\ \ ratios\ \ at
\DAFNE}

The semileptonic branching ratios of the $K_S$ can be measured at
\DAFNE\ by selecting the following final states
$|l^\pm \pi^\mp \nu(t_1),x_L(t_2)\rangle$, where
 $t_1 \leq 10 \tau_S$, $t_2 \geq 10 \tau_S$ and $|x_L\rangle$ is one
of the allowed final states in $K_L$ decays
($|\pi^+\pi^-\pi^0\rangle$ or $|l^\pm \pi^\mp \nu\rangle$).
The probability of such events is obtained by
integrating  eq. (\ref{gamab})\footnote{
If CPT is not conserved the only change in time
evolution equations concerns the normalization factor, which becomes $h =
\sqrt{(1+ |\epsilon_L|^2)(1 +|\epsilon_S|^2)}/(1 -\epsilon_L \epsilon_S)
\simeq 1$.}    in the appropriate time intervals\footnote{
The general constraints on the time intervals for $K_S$ tagging are:
$t_1 \leq t_1^{\rm max}$ and $t_2 \geq t_2^{\rm min}$, with
$\tau_S \ll t_1^{\rm max} \leq t_2^{\rm min} \ll \tau_L$. A good
choice is given by $ t_1^{\rm max}= t_2^{\rm min}=10\tau_S$. }.
Therefore the number of events for
$N_0$ initial $K_S K_L$ pairs is given by:
\begin{eqnarray}
N_S(l^\pm) = N_0\Big\lbrace {\rm Br}(K_S \to l^\pm \pi^\mp \nu)
{\rm Br}(K_L \to x_L)
S_1 +  {\rm Br}(K_L \to l^\pm \pi^\mp \nu) {\rm Br}(K_S \to x_L) S_2
\nonumber \\
- \Re\Big[ S_3\int A(K_S \to l^\pm \pi^\mp \nu)
 A^*(K_L \to l^\pm \pi^\mp \nu)
 A(K_L \to x_L)A^*(K_S \to x_L) {
d\phi_{\pi l \nu}d\phi_{x_L} \over \Gamma_S \Gamma_L} \Big]\Big\rbrace,
\nonumber \\
\label{Nslpm}
\end{eqnarray}
where
\begin{eqnarray}
 S_1 =&\left(1-e^{-10}\right)
 \left(e^{-10{\Gamma_L \over \Gamma_S} } - e^{-{D \over d_L}}\right)&=
0.28 \simeq S_L \nonumber\\
 S_2 =&  \left(1 -e^{-10{\Gamma_L \over \Gamma_S} }\right) e^{-10}
&= 7.8\times10^{-7} \label{sfac} \\
 S_3 =& \displaystyle\frac{ 2\Gamma_S \Gamma_L}{| \Gamma|^2 +| \Delta m|^2}
e^{-10{(\Gamma +i \Delta m)\over \Gamma_S}} &= (0.3 - i 4.8)\times 10^{-5}.
\nonumber
\end{eqnarray}
As can be seen, $S_1$ is by far the dominant contribution; the branching
ratio products in eq. (\ref{Nslpm})
are predicted to be of the same order, while the interference
term should be further suppressed by large cancellations.
Therefore inserting the experimental value \cite{PDG}
${\rm Br}(K_L \to x_L) = (78.1 \pm 0.7 )\%$ eq. (\ref{Nslpm})  becomes:
\begin{equation}
N_S(l^\pm) = 0.22 \times N_0 \times {\rm Br}(K_S \to l^\pm \pi^\mp \nu).
\label{nslpmb}
\end{equation}
The project luminosity of \DAFNE\ (${\cal L} = 5\times 10^{32}\ {\rm cm}^2
{\rm s}^{-1}$)
gives about $8.6\times10^9 K_L K_S$/year.
Using eq. (\ref{gsgl}) and the present upper limit on the
violation of the $ \Delta S = \Delta Q$  rule \cite{PDG}, we estimate
${\rm Br}(K_S \to \mu^\pm \pi^\mp \nu) = 4.66\times 10^{-4}$
and ${\rm Br}(K_S \to e^\pm \pi^\mp \nu)= 6.68\times 10^{-4}$,
therefore  $2.1\times 10^6$
events/year are expected.\par
With these numbers we can estimate the sensitivity of \DAFNE\ to CP, CPT
and the ${\Delta S=\Delta Q}$ rule violating parameters defined in
eqs. (\ref{deltasl}) and (\ref{eta}).
Since the $\mu^+ \pi^- \nu$  final state
can hardly be distinguished
from the $\mu^- \pi^+ \bar{ \nu}$ one,  we conservatively assume that
only electrons can be used to derive $K_S$ charge asymmetry.
In this case the number of event is $1.2 \times 10^6$/year
and the statistical error
on $\delta_S$ turns out to be $\sigma_{\delta_S}= 9.0\times 10^{-4}$.
Since the experimental value of $K_L$ charge asymmetry is
$\delta_L = (3.27 \pm 0.12)\times 10^{-3}$
\cite{PDG}, we expect $\sigma_{\delta_S-\delta_L}/\delta_L\simeq 0.28$,
testing the CPT prediction $\delta_S=\delta_L$
at a significant level.\par
Eq. (\ref{eta}) (test of $\Delta S =\Delta Q$ rule) involves
the semileptonic rates of $K_S$ and $K_L$;
thus to estimate the error on $\eta$
one has to take into account also the experimental errors on tagging
branching ratios and on  $K_{S,L}$ widths. Using the values in
Ref. \cite{PDG}, these effects give a large contribution to the total error,
which turns out to be $\sigma_{\eta}=1.1\times 10^{-2}$, whereas
the pure statistical contribution would give only $=1.4\times 10^{-3}$.
This large value for $\sigma_{\eta}$ will perhaps  be lowered by
measuring all the quantities involved in the same experimental set-up.
\par
In Table 1 we report  the predicted sensitivity of \DAFNE\ in
comparison with other experiments.
As one can see, \DAFNE\
is very powerful to test $\Delta S = \Delta Q$ rule.

\vskip .5 true cm
\begin{eqnarray}
  \begin{array}{|c|c|c|c|c|} \hline
{\rm Parameter}  & {\rm PDG}   & \multicolumn{2}{|c|}{\rm CPLEAR}
   & {\rm DA}\Phi{\rm NE} \\
   &  & \sigma('93) & \sigma('95) & \sigma(1~{\rm yr}) \\ \hline
\ \ \delta_L\hfill & (3.27\pm 0.12)\times 10^{-3} & \multicolumn{2}{|c|}{-} &
   0.04 \times 10^{-3} \\ \hline
\ \ \delta_S\hfill & - & \multicolumn{2}{|c|}{-}& 0.9\times 10^{-3} \\ \hline
\ \ \Re\epsilon_S\hfill & - & 0.7\times 10^{-3} & 0.4\times 10^{-3}
& - \\ \hline
\ \ \Re (c^*/a)=\Re x\hfill & (6\pm18)\times 10^{-3} & 8\times 10^{-3}
   & 5\times 10^{-3} & 3 \times 10^{-3} \\ \hline
\ \ A_T\hfill & - & 2\times 10^{-3} & 1\times 10^{-3} &
   2 \times 10^{-3} \\ \hline
\ \ A_{CPT}\hfill & - & 2\times 10^{-3} & 1\times 10^{-3} &
   2 \times 10^{-3} \\ \hline
\end{array}
\nonumber
\end{eqnarray}
\noindent
Table 1: Comparison between the present experimental data \cite{PDG},
CPLEAR present and expected sensitivity \cite{CPLEAR}~ and the
achievable sensitivity in 1 year at DA$\Phi$NE, for the semileptonic
parameters. For both CPLEAR and DA$\Phi$NE only the statistical
error has been reported. Note that  $A_T$ and $A_{CPT}$ asymmetries
have different theoretical expressions, for
CPLEAR and DA$\Phi$NE, if one considers CPT
violation in the decay amplitudes.

\subsection{Direct tests of T and CPT symmetries}

The dilepton events allow direct tests of T and CPT symmetries
\cite{ohs,bucha}.
A long time ago Kabir \cite{kabir} showed that T violation implies
different probabilities for $K^0\rightarrow \bar{K^0}$ and
$\bar{K^0}\rightarrow K^0$ transitions, while CPT requires equal probabilities
for  $K^0\rightarrow K^0$ and  $\bar{K^0}\rightarrow \bar{K^0}$ transitions.
Then a T-violating asymmetry:
\begin{equation}
A_T =  \frac {   P \left( \bar{K^0} \rightarrow K^0 \right)-
                      P \left( K^0 \rightarrow \bar{K^0} \right)  }
                  {   P \left( \bar{K^0} \rightarrow K^0 \right) +
                      P \left( K^0 \rightarrow \bar{K^0} \right) }
\label{asymt}
\end{equation}
and a CPT-violating one:
\begin{equation}
A_{CPT} = \frac {  P \left( K^0 \rightarrow K^0 \right) -
                        P \left( \bar{K^0} \rightarrow \bar{K^0} \right)
                        }
                     {  P \left( \bar{K^0}\rightarrow \bar{K^0} \right) +
                        P \left( K^0 \rightarrow K^0 \right) }
\label{asymcpt}
\end{equation}
can be defined.
\par
Both these tests can be done at a $\phi$ factory, where the initial state
is an antisymmetric $K^0 \bar{K^0}$ state, if the $\Delta S =\Delta Q$ rule
holds.

If a neutral kaon decays into a positive lepton at a time $t$, the
other neutral kaon is at the same time a $\bar{K^0}$ and
 the sign of the lepton, emitted
in a subsequent semileptonic decay, signals if the $\bar{K^0}$ has changed or
conserved its own flavour. Therefore,
if ${|A(K^0 \to l^+ x)|^2} \
= \ {| A(\bar K^0 \to l^- x)|^2}$, the charge asymmetry
in equal-sign dilepton pairs measured at the $\phi$ factory will be
 equal to $A_T$. On the other hand, time asymmetry in opposite-sign
dilepton pairs signals CPT violation.

In the more general case, taking into account
also possible violations of the $\Delta S = \Delta Q$ rule
one gets\footnote{In the following
equations we include also the CPT-conserving higher-order terms,
namely the ones proportional to $\epsilon_M{ c^*\over a}$.}:
\begin{equation}
A_T = \frac{ L^{++} - L^{--} }{ L^{++} + L^{--} }
         = 2\Re\left( \epsilon_L + \epsilon_S \right) +
            4\Re\left(\frac{b}{a}\right) = \delta_L + \delta_S,
\label{asymtn}
\end{equation}
and
\begin{eqnarray}
 A_{CPT} &=&
        \frac{L^{-+}-L^{+-}}{L^{-+}+L^{+-}} =
        2 \Re \left( \epsilon_L -  \epsilon_S \right)  +
        4 \Re \left(\frac{d^*}{a}\right)  +
        4 \Re\left(\epsilon_L + \epsilon_S\right)
              \Re \left(\frac{c^*}{a}\right) \nonumber  \\
        & &
    + \frac{4}{S_L}\left[ \Im\left(\epsilon_L - \epsilon_S\right) -
      2\Im\left(\frac{c^*}{a}\right)+2\Im\left(\epsilon_L + \epsilon_S\right)
     \Re\left(\frac{c^*}{a}\right) \right]
        \frac{\Delta m \Gamma_L}{\Gamma^2 + \Delta m^2},
\label{asymcptn}
\end{eqnarray}
where $L^{+-} \left( L^{-+} \right)$ is the number of dilepton pairs
with the positive lepton  emitted before (after) the negative one.
The number of equal-sign electron pairs ($L^{++} + L^{--}$)
and that of opposite-sign ($L^{+-} + L^{-+}$) expected
at \DAFNE\ is about $3.3 \times10^5$ events/year,
therefore the T- and CPT-violating asymmetries can be measured with a
statistical error of about
$1.7 \times 10^{-3}$.
\par
Violation of the $\Delta S = \Delta Q$ rule does not affect eq. (\ref{asymtn})
but the CPT violation in the decay amplitude contributes together with the
true T-violating term $( \epsilon_L + \epsilon_S )$.
On the contrary in eq. (\ref{asymcptn})
the effects of CPT violation and $\Delta S = - \Delta Q$
transitions cannot be disentangled.
In the  CPT limit the time asymmetry can be written as:
\begin{equation}
\bar{A}_{CPT} = 8 \Re(\epsilon_M) \Re\left({c^*\over a}\right) -
8 \frac{ \Im \left({c^*\over a} \right)  -
2\Im\epsilon_M \Re\left(\frac{c^*}{a}\right)}
{S_L} \frac{\Delta m \Gamma_L}{\Gamma^2 + \Delta m^2}
\end{equation}
and inserting the experimental limits on $c^*/a$ \cite{PDG} one has:
\begin{equation}
 | \bar{A}_{CPT}| < 1.1 \times 10^{-3}.
\end{equation}
Thus a value of $A_{CPT}$ larger than $ 10^{-3}$ indicates an actual
CPT violation either in the kaon mass matrix or in $ \Delta S = - \Delta Q$
transition amplitudes. \par

More information can be obtained by
the study of the time dependence of opposite sign dilepton events.
Choosing for the final states of eq. (\ref{intt})
$|a\rangle = |e^+ \pi^- \nu\rangle$
and $|b\rangle = |e^- \pi^+ \bar{\nu} \rangle$ and integrating over the
phase space one gets:
\begin{eqnarray}
 && I(e^+,e^-;t) ={\Gamma_L^e\Gamma_S^e \over 8\Gamma} e^{-\Gamma|t|}
\left\lbrace \left[ 1+4\Re\Delta -4\Re\left({d^*\over a}\right)
-8\Re\epsilon_M\Re\left({c^*\over a}\right)\right]
e^{-{\Delta\Gamma\over 2}t}  \right. \nonumber \\
&&\hspace{2.9 cm} +\left[ 1-4\Re\Delta +4\Re\left({d^*\over a}\right)
+8\Re\epsilon_M\Re\left({c^*\over a}\right)\right]
e^{+{\Delta\Gamma\over 2}t} +2\cos(\Delta m t) \nonumber \\
&&\hspace{2.9 cm}  \left. -8\left[ \Im\Delta +\Im\left({c^*\over a}\right)
-2\Im\epsilon_M  \Re\left(\frac{c^*}{a}\right)\right]
\sin(\Delta m t)
\right\rbrace  \nonumber \\
&& = {\Gamma_L^e\Gamma_S^e \over 4\Gamma} e^{-\Gamma|t|}
\left[\cosh \left(\frac{ \Delta \Gamma t}{2}\right) + \cos( \Delta m t)
- 4 \Delta_R\sinh \left(\frac{ \Delta \Gamma t}{2}\right) - 4 \Delta_I
\sin( \Delta m t) \right]
\label{intlplm}
\end{eqnarray}
where
\begin{equation}
\Delta_R =\Re\Delta -\Re\left({d^*\over a}\right)
-2\Re\epsilon_M\Re\left({c^*\over a}\right) , \;\;
\Delta_I = \Im\Delta +\Im\left({c^*\over a}\right)
-2\Im\epsilon_M  \Re\left(\frac{c^*}{a}\right).
\end{equation}
The difference in the asymptotic limits ($ |t| \gg \tau_S $) leads
to the determination of $\Delta_R$,
while the interference term singles out
$\Delta_I$.
The higher-order terms can be neglected (their upper bound is about
$0.6\times 10^{-4}$, smaller than the \DAFNE\ sensitivity),
but the CPT-violating parameter $\Delta$ and the $ \Delta S =
- \Delta Q$ contributions are still mixed.
An exact determination of the statistical error on  $\Delta_R$ and  $\Delta_I$
would require a simulation of the experimental apparatus, which is
beyond the purpose of this work. To give an idea of the \DAFNE~ sensitivity
we report in Fig. 3 the asymmetry in opposite-sign dileptons as a function
of the time difference
\begin{equation}
A_{CPT}(t) = \frac{ I(e^+,e^-;|t|)- I(e^+,e^-;-|t|)}{ I(e^+,e^-;|t|)+
I(e^+,e^-;-|t|)},
\end{equation}
for $\Delta_I =0$ and  $\Delta_I =\pm 5 \Delta_R$.
As  can be seen the asymptotic value is reached very soon and the three
curves are clearly distinct. Therefore we estimate $ \sigma_{\Delta_R}
\simeq \sigma_{A_{CPT}}/4 \simeq 5 \times 10^{-4}$. The value of
 $\sigma_{\Delta_I}$ depends critically on the experimental resolution.
We estimate that, as   happens for the real and the imaginary parts  of \rap,
$\sigma_{\Delta_I}$ will be about 20 times larger than  $\sigma_{\Delta_R}$.

 \begin{figure}[t]  
     \begin{center}
        \setlength{\unitlength}{1truecm}
        \begin{picture}(8.0,8.0)
           \put(-4.0,-6.0){\special{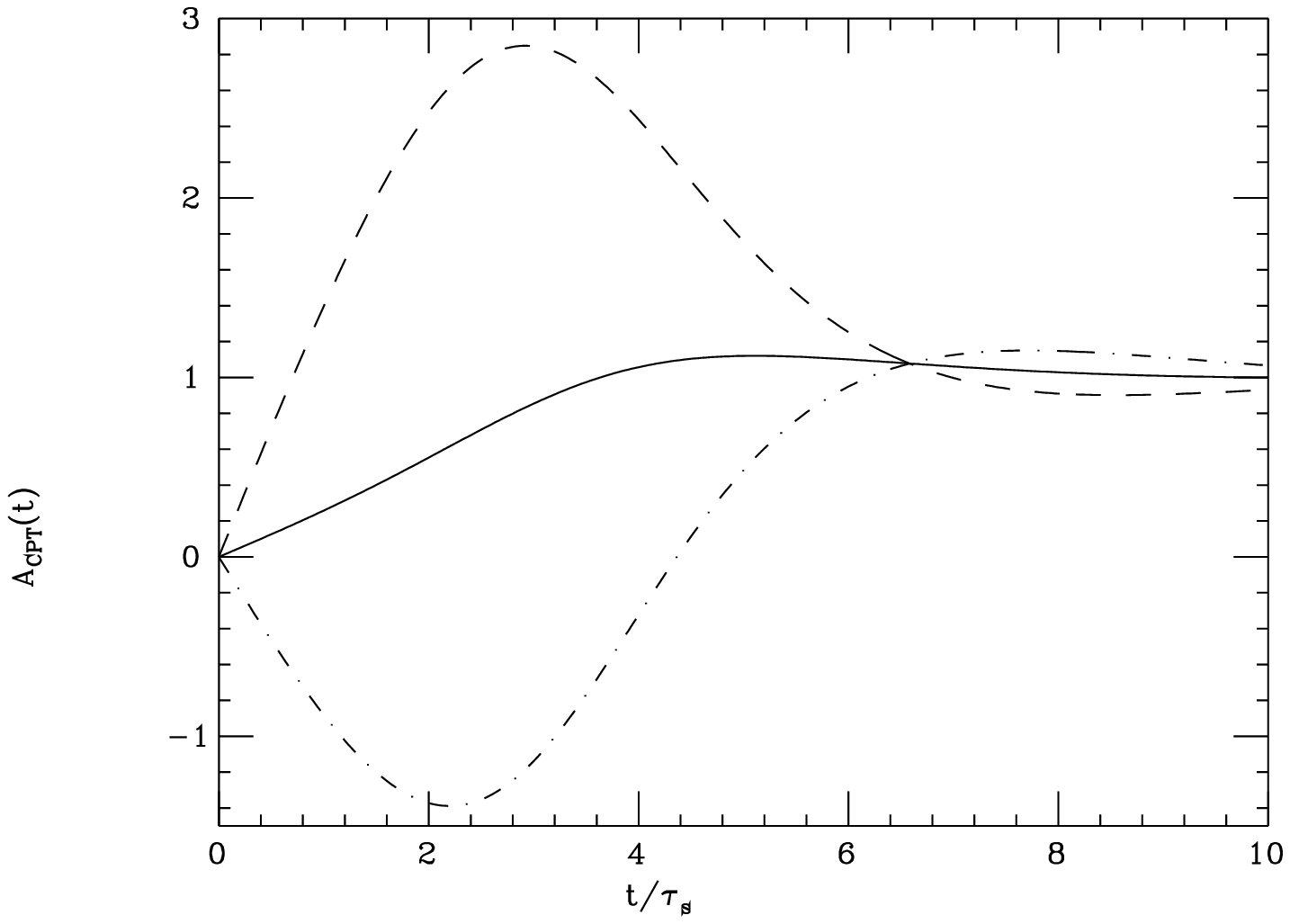}}
        \end{picture}
    \end{center}
    \caption{The asymmetry $A_{CPT}(t)$ as a function of the time difference
for $\Delta_R = 1/4$. The full, dashed and dot-dashed lines correspond to
$\Delta_I = 0$, $\Delta_I = 5\Delta_R$ and $\Delta_I = -5\Delta_R$
respectively.}
    \protect\label{figIep}
\end{figure}

As shown in Ref. \cite{maiani}, the inclusion in the analysis of
the $K \to \pi \pi$ decays allows us  to
disentangle almost all the amplitudes. Indeed,
in the Wu-Yang phase convention,\footnote{
The $\pi\pi$ decay amplitudes are parametrized in each
isospin channel like the semileptonic ones, $A_I$ is the CPT-conserving
part and $B_I$ the CPT-violating one. The Wu-Yang convention is
$\Im(A_0)=0$.} unitarity implies that
the phase of $\epsilon_M$ is equal to
$\phi_{SW}\equiv \arctan(2 \Delta m / \Delta \Gamma) = (43.64 \pm
0.15)^{\circ}$
and the phase of $\left[\Delta - \Re(B_0)/A_0 \right]$ is
$\left[\phi_{SW} \pm {\pi \over  2}\right]$;
therefore, one has \cite{bucha,maiani}:
\begin{equation}
 \frac{1}{3} (2\eta^{+-} + \eta ^{00}) = |\epsilon_M|e^{i\phi_{SW}}
\pm i| \Delta - \frac{\Re B_0}{A_0}|e^{i\phi_{SW}}.
\end{equation}
The present experimental data on $\eta^{+-}$ and $\eta ^{00}$ \cite{PDG}
give:
\begin{equation}
  |\Delta - \frac{\Re B_0}{A_0} | =
   (1.7 \pm 2.8) \times 10^{-5} \qquad {\rm and} \qquad
  |\epsilon_M| =
   (2.266 \pm 0.017) \times 10^{-3}.
\end{equation}
 As  can be seen, the CPT constraint is at present very well satisfied
and, assuming CPT conservation in decay amplitudes, the limit in $K^0
\bar{K}^0$ mass difference is
 \begin{equation}
{|M_{11} - M_{22}|\over m_K} \leq  10^{-18} \ {\rm GeV},
 \end{equation}
close to the natural scale factor $ m_K /M_{\rm Planck}$. \par
In addition, from the measured value of $K_L$ charge asymmetry one gets:
\begin{equation}
   -\Re \Delta + \Re\left( \frac{b}{a}\right) +\Re\left(\frac{d^*}{a}\right)
   = (-0.06 \pm 0.6) \times
   10^{-4}.
\end{equation}
The future measurement  of $\delta_S$ at \DAFNE~ would lead also
to the determination of $\Re (b/a)$, while the
CPLEAR experiment will give direct measurement
of $\Re \epsilon_S$ and of $\Im (c^*/a)$.
Therefore all the parameters that appear in the observables introduced above
can be disentangled, and some consistency relations must be satisfied.
The preliminary data of the CPLEAR collaboration \cite{CPLEAR}
have  large errors and still do not give significant bounds.
We report in Table 2 the relations between the observables and the
theoretical parameters with the corresponding statistical errors
from present and future experiments, together with
the consistency equations and the corresponding sensitivity.
To simplify the notations of the table  we define
\begin{equation}
w=\Delta - \frac{\Re B_0}{A_0} = \epsilon_M -
\frac{1}{3} (2\eta^{+-} + \eta ^{00}) .
\end{equation}
\vskip .5 true cm
\begin{eqnarray}
  \begin{array}{|c|c|c|} \hline
{\rm Equations}  & {\rm Parameters}   & \sigma   \\ \hline
\ \ \Re \Delta = \Re \epsilon_S - \Re \epsilon_M \hfill
    & \Re \Delta & 4 \times 10^{-4}   \\ \hline
\ \ \frac{\Re B_0}{A_0}= \Re \epsilon_S - \Re\epsilon_M - \Re w \hfill
    & \frac{\Re B_0}{A_0} & 4 \times 10^{-4} \\ \hline
\ \ \Re(\frac{b}{a})=\frac{1}{4}(\delta_L+\delta_S) -\Re \epsilon_M  \hfill
    & \Re(\frac{b}{a}) & 2 \times 10^{-4} \\ \hline
\ \ \Re(\frac{d^*}{a})=\Re\epsilon_S-\Re \epsilon_M+\frac{1}{4}(\delta_L
    -\delta_S)\hfill & \Re (\frac{d^*}{a}) & 5 \times 10^{-4} \\ \hline
\ \ {\frac{1}{4}(\delta_L+\delta_S)-\frac{1}{4}A_{T}=0}\hfill & -
    & 6\times 10^{-4} \\ \hline
\ \ {\frac{1}{4}(\delta_L-\delta_S)+\Delta_R =0 }\hfill & -
    & 6\times 10^{-4} \\ \hline
\ \ {\Im ({c^*\over a})+ \Im w -  \Delta_I =0 }\hfill & -
    & \sim 10^{-2} \\ \hline
\end{array}
\nonumber
\end{eqnarray}
\noindent
Table 2: Statistical errors on parameters and consistency relations, using
present experimental data \cite{PDG} (for $\eta^{+-}$, $\eta^{00}$
 and $\delta_L$) together with
\DAFNE~ (for  $\delta_S$, $A_T$, $\Delta_R$ and $\Delta_I$)
and CPLEAR (for $\Re\epsilon_S$ and $\Im(c^*/a)$) future results.
The $\sigma$ of the last equation in the table  is only a guess.

\section{$K_S\to 3\pi$}

The $K_S\to 3\pi^0$ decay is a pure CP-violating transition,
while the $K_S\to \pi^+\pi^-\pi^0$ decay receives both CP-conserving and
CP-violating contributions.

The CP-conserving $K \to 3 \pi$ decay amplitudes are well described by
Chiral Perturbation Theory (ChPT). They have been calculated, including
 the  next-to-leading-order corrections, in Ref. \cite{KMW} and turn out
to be in good agreement with the experimental data.
The CP-conserving $K_S\to \pi^+\pi^-\pi^0$ decay amplitude is odd under
$\pi^+ -\pi^-$  momenta exchange and thus, neglecting final states
with high angular momenta, it is induced by
a $\Delta I=3/2$ transition. The ChPT  calculation of Ref. \cite{KMW}
leads to the  prediction:
\begin{equation}
{\rm Br}(K_S\to \pi^+\pi^-\pi^0)_{CP^+}= (2.4\pm 0.7) \times 10^{-7},
\label{brs1}
\end{equation}
consistent  with the recent data:
\begin{eqnarray}
{\rm Br}(K_S\to \pi^+\pi^-\pi^0)_{CP^+}&=(3.9^{+5.4\ +0.9}_{-1.8\ -0.7})
\times 10^{-7},	\hspace{1.5 cm}&{\rm E621}\cite{thomson} \\
{\rm Br}(K_S\to \pi^+\pi^-\pi^0)_{CP^+}&=(7.8^{+5.7\ +7.3}_{-4.1\ -4.9})
\times 10^{-7}.	\hspace{1.5 cm}&{\rm CPLEAR}\cite{CPLEAR}
\end{eqnarray}

As in  $K\to 2\pi$ decays, for the CP-violating amplitudes
it is convenient to define the ratios:
\begin{eqnarray}
\eta^{+-0} &=&
\displaystyle{(A_S^{+-0})_{CP^-}\over A_L^{+-0}} =\epsilon_S
+\epsilon'_{+-0},\\
\eta^{000} &=&
\displaystyle{A_S^{000} \over A_L^{000}} =\epsilon_S +\epsilon'_{000}.
\end{eqnarray}
The direct CP-violating parameters $\epsilon'_{+-0}$ and $\epsilon'_{000}$
have been evaluated at lowest order in ChPT \cite{liwo} and  turn out to
be of the same order as $\epsilon'$. As shown in  \cite{dong},
higher-order terms can substantially enhance
$\epsilon'_{+-0}$ and $\epsilon'_{000}$, which are nevertheless
negligible\footnote{
In $K\to 3\pi$ we will neglect possible CPT-violating effects, thus in the
following we will assume $\epsilon_S=\epsilon_L=\epsilon$.}
compared  to $\epsilon$. The
predicted branching ratios are:
\begin{eqnarray}
&{\rm Br}\left(K_S\to (\pi^+\pi^-\pi^0)_{CP^-}\right)&\simeq
1.06\times 10^{-9},\\
&{\rm Br}(K_S\to 3\pi^0)  &\simeq 1.94\times 10^{-9},
\label{brs2}
\end{eqnarray}
much smaller then the present upper limits \cite{PDG, CPLEAR, thomson2}.

Due to the smallness of the branching ratios
it is very hard to detect  $K_S\to 3 \pi$ decays, especially the
CP-violating ones. Tagging the $K_S$ as in the case of the semileptonic
decays (eqs. (\ref{Nslpm}) and (\ref{sfac}))
and inserting the numerical values,
one gets for $3\pi^0$ final state:
\begin{eqnarray}
N_S(3\pi^0) = N_0{\rm Br}(K_L \to 3 \pi^0) [
3.8 \times 10^{-4}|\eta^{000}|^2
 + 8.9 \times 10^{-10} - 1.8 \times 10^{-10}
\Im(\eta^{000})].
\label{ksppp}
\end{eqnarray}
The total number of events is very small ($\simeq $ 6 per year)
and the ratio of right events (those with a $K_S\to 3 \pi^0$ decay)
to wrong ones (those with a $K_L\to 3 \pi^0$ decay)
is only 2.2.

In the case of the CP-conserving $K_S\to \pi^+\pi^-\pi^0$ decay,
the expected number of events is about 440 with a negligible
background.

A more promising way to detect the CP-violating
$K_S\to 3\pi$ decays is to study the
interference terms of $I(a,b;t)$, in eq. (\ref{intt}), choosing
$|a\rangle=|3\pi\rangle$ and $|b\rangle=|l^\pm \pi^\mp \nu\rangle$,
as suggested in Refs. \cite{NG,noi}.
For the $K_S\to 3\pi^0$ it is useful to define the asymmetry:
\begin{equation}
A^{000}(t) = \displaystyle{  \int\left[ I(3\pi^0, l^+ \pi^- \nu; t) -
 I(3\pi^0, l^- \pi^+ \nu; t) \right]d\phi_{3\pi} \ d\phi_{l\pi\nu}
\over \int\left[ I(3\pi^0, l^+ \pi^- \nu; t) +
 I(3\pi^0, l^- \pi^+ \nu; t) \right]d\phi_{3\pi} \ d\phi_{l\pi\nu}}
\end{equation}
which, integrating over the $3\pi$ and $\pi l\nu$ Dalitz plots,
becomes:
\begin{equation}
A^{000}(t) = \displaystyle{ 2 \Re \epsilon e^{+{\Delta \Gamma  \over 2}t}
-2\Re\left(\eta^{000}e^{+i \Delta m t}\right)  \over
 e^{+{\Delta \Gamma  \over 2}t} +|\eta^{000}|^2
 e^{-{\Delta \Gamma  \over 2}t}  }.
\label{a000}
\end{equation}
For positive and large values of the time difference $t$, eq. (\ref{a000})
reads: $A^{000} \simeq 2\Re \epsilon$;
on the other hand, for negative value of $t$, one gets an interesting
interference effect between $\epsilon$ and $\eta^{000}$, as shown in Fig. 4.
The asymmetry for $t<0$ is quite large, but the total number of events
is small, about $2\times 10^3$ per year.

 \begin{figure}[t]  
     \begin{center}
        \setlength{\unitlength}{1truecm}
        \begin{picture}(8.0,8.0)
           \put(-4.0,-6.0){\special{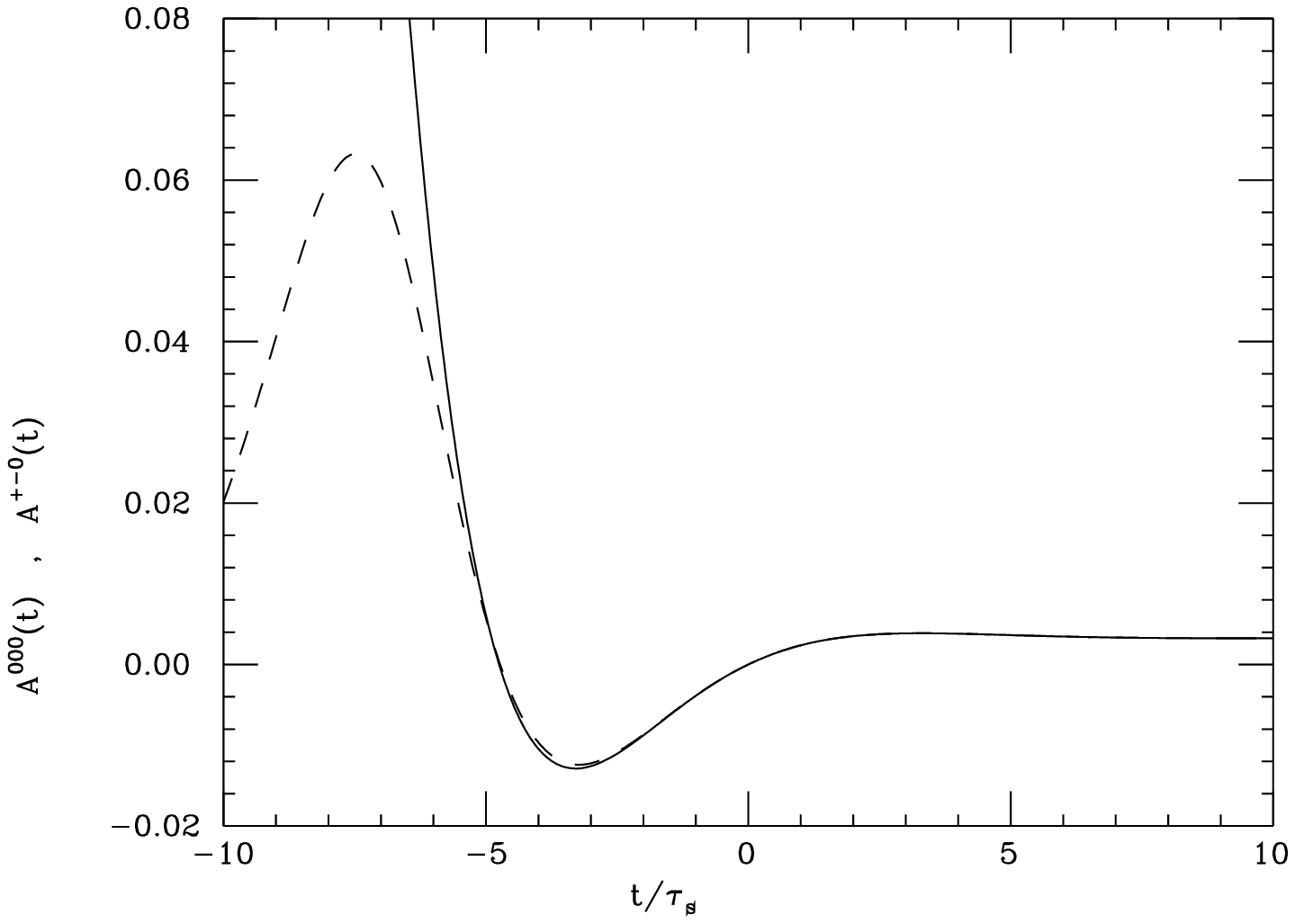}}
        \end{picture}
    \end{center}
    \caption{The asymmetries $A^{000}(t)$ (full line) and $A^{+-0}(t)$
(dashed line). We have fixed  $ \eta^{000} =\eta^{+-0}=
|\epsilon|e^{i\phi_{SW}}$.}
    \protect\label{figksa}
\end{figure}

In the case of the $\pi^+\pi^-\pi^0$ final state, the CP-violating and
CP-conserving amplitudes have opposite symmetry under
$\pi^+ -\pi^-$  momentum exchange. Therefore it is possible to select
the CP-violating and the CP-conserving part of the interference term in
eq.(\ref{intt}) with an even or an odd integration over the
$3\pi$ Dalitz plot. Analogously to the $3\pi^0$ case,
for the CP-violating part we define the asymmetry:

\begin{eqnarray}
A^{+-0}(t)
&=& \displaystyle{  \int\left[ I(\pi^+\pi^-\pi^0, l^+ \pi^- \nu; t) -
 I(\pi^+\pi^-\pi^0, l^- \pi^+ \nu; t) \right]d\phi_{3\pi}\, d\phi_{l\pi\nu}
\over \int\left[ I(\pi^+\pi^-\pi^0, l^+ \pi^- \nu; t) +
 I(\pi^+\pi^-\pi^0, l^- \pi^+ \nu; t) \right]d\phi_{3\pi} \, d\phi_{l\pi\nu}},
 \nonumber \\
        &=& \displaystyle{ 2(\Re \epsilon)e^{+{\Delta \Gamma  \over 2}t}
-2\Re\left(\eta^{+-0}e^{+i \Delta m t}\right)  \over
 e^{+{\Delta \Gamma  \over 2}t} +{\Gamma_S^{+-0} \over \Gamma_L^{+-0}}
 e^{-{\Delta \Gamma  \over 2}t}  },
\label{apm0}
\end{eqnarray}
while the CP-conserving part can be singled out by the ratio:
\begin{eqnarray}
R^{\pm}(t)
&=& \displaystyle{  \int_+
 I(\pi^+\pi^-\pi^0, l^\pm \pi^\mp \nu; t)d\phi_{3\pi} \, d\phi_{l\pi\nu} -
 \int_-
 I(\pi^+\pi^-\pi^0, l^\pm \pi^\mp \nu; t)d\phi_{3\pi} \, d\phi_{l\pi\nu}
\over \int\left[ I(\pi^+\pi^-\pi^0, l^\pm \pi^\mp \nu; t)
 \right]d\phi_{3\pi} \, d\phi_{l\pi\nu}},
 \nonumber \\
&=& \pm 2\displaystyle{ \int_+ \Re\left(A^{+-0}_L A^{+-0*}_S \right)
d\phi_{3\pi} \over
 \Gamma_L^{+-0} e^{+{\Delta \Gamma  \over 2}t} +\Gamma_S^{+-0}
e^{-{\Delta \Gamma  \over 2}t} }
\left[\cos(\Delta mt)+{\widetilde \delta}\sin(\Delta mt) \right],
\label{Rp}
\end{eqnarray}
where $\int_{\pm} d\phi_{3\pi}$ indicates the integration in
the region $E_{\pi^\pm} > E_{\pi^\mp}$.

 \begin{figure}[t]  
     \begin{center}
        \setlength{\unitlength}{1truecm}
        \begin{picture}(8.0,8.0)
           \put(-4.0,-6.0){\special{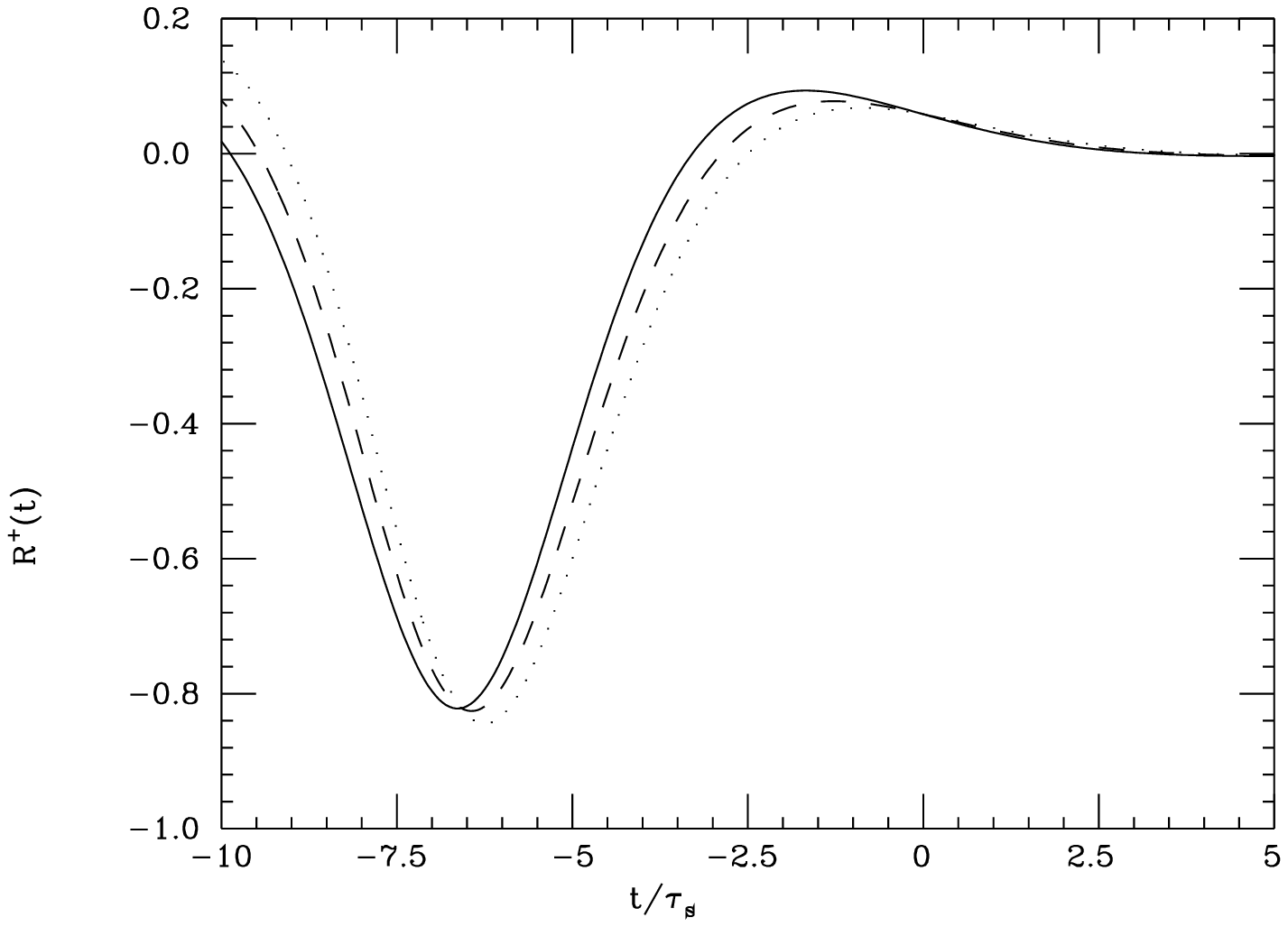}}
        \end{picture}
    \end{center}
    \caption{The ratio $ R^+(t)$ defined in eq.(\protect\ref{Rp}).
 The full, dashed and dotted
lines correspond to ${\widetilde \delta} = 0$, ${\widetilde \delta}= 0.2$ and
${\widetilde \delta} = 0.4$ respectively.  }
    \protect\label{figksp}
\end{figure}

The behaviour of $A^{+-0}(t)$ is completely analogous to the one of
$A^{000}(t)$ (Fig. 4). As discussed in refs. \cite{NG,noi} the study of
$R^\pm(t)$ will certainly lead to a determination of the
$(A_S^{+-0})_{CP^+}$ amplitude, performing an interesting test of
ChPT in the $\Delta I=3/2$ transitions, and perhaps could also lead
to a direct measurement of the $\pi^+\pi^-\pi^0$ rescattering functions.
The phase ${\widetilde \delta}$ of eq. (\ref{Rp}) can be written as:
\begin{equation}
{\widetilde \delta}\simeq\delta_{1S}-\delta_{2},
\end{equation}
where $\delta_{2}$ and $\delta_{1S}$ are the $3\pi$ strong phases
of the $I=2$ and of the symmetric $I=1$ final states, respectively.
The ChPT prediction is ${\widetilde \delta}=(10\pm 1)^\circ$
\cite{noi} and the first measurement \cite{thomson} gives
${\widetilde \delta}=(59\pm 48)^\circ$.
With a different integration over the Dalitz plot,   also the phase of
the non-symmetric $I=1$ final state could be selected \cite{NG,noi}.
Figure 5 shows the behaviour of $R^\pm$ for different values of
${\widetilde \delta}$.

\section{Interference in $K\to\pi\pi\gamma$}
\label{sec:kpipigamma}
In this section we will discuss the possibility to detect CP
violation through the study of the time difference distribution  $I(a,b;t)$,
defined in eq. (\ref{intt}), choosing $|a\rangle=|\pi^+ \pi^- \gamma\rangle$
 and $|b\rangle=|\pi^\pm l^\mp \nu\rangle$ \cite{NG}.
To this purpose we recall some useful
decomposition  of the $K\to\pi^+\pi^-\gamma$ decay amplitude, referring to
\cite{dibP} for a more detailed  discussion of these decays.

The amplitude for $K(p_{_K})\rightarrow\pi^+(p_+)
\pi^-(p_-)\gamma(q,\varepsilon)$ decays can be generally decomposed as
the sum of two terms: the inner bremsstrahlung $(A_{IB})$ and
the direct emission $(A_{DE})$ \cite{dibP}.
The first term, which has a pole at zero photon energy,
is completely predicted by QED in terms of the  $K\to\pi^+\pi^-$ amplitude
\cite{Low}:
\begin{equation}
A_{IB}( K_{S,L}\to\pi^+\pi^-\gamma) =e\left({\varepsilon \cdot p_- \over q
\cdot p_-}-{\varepsilon \cdot  p_+ \over q \cdot p_+ } \right) A(
K_{S,L}\to\pi^+\pi^-).
\label{defib}
\end{equation}
The second term, which is obtained by
subtracting $A_{IB}$ from the total amplitude,
depends on the structure of the
$K\to\pi^+\pi^-\gamma$ effective vertex
and provides a test for mesonic interaction models.

The $K \rightarrow \pi^+\pi^-\gamma$ amplitude
is usually decomposed also in a different way, separating the
electric and the magnetic terms. Defining
the dimensionless amplitudes $E$ and $M$ as in \cite{ENPb},
we can write:
\begin{equation}
A(K_{S,L}\to\pi^+\pi^-\gamma)=
\varepsilon_\mu(q)\left[E(z_i)(p_+ \cdot q \,p_-^{\mu}
-p_- \cdot q \, p_+^{\mu})+M(z_i)\epsilon^{\mu\nu\rho\sigma}p_{+\nu} p_{-\rho}
q_\sigma \right] /m_K^3,   \label{amplitude}
\end{equation}
where
\begin{equation}
z_\pm={p_\pm\cdot  q \over m_K^2}, \qquad\qquad{\rm and }\qquad \qquad
z_3={p_{_K} \cdot q \over m_K^2}=z_++z_-.
\end{equation}
As   can be seen from eq. (\ref{defib}),
the inner bremsstrahlung amplitude can contribute only to the
$E(z_i)$ term, while the direct emission amplitude can
contribute to both  the $E(z_i)$ and the $M(z_i)$ terms.
Summing over photon helicities there  is no interference between electric and
magnetic terms:
\begin{equation}
{\rm d} \Gamma =
{m_K\over{(4\pi)^3}} (|(E(z_i)|^2+|M(z_i)|^2)\left[
z_+z_-(1-2z_3- 2r_m^2)-r_m^2(z_+^2 +z_-^2)\right] {\rm d}z_+ {\rm d}z_-
\label{width}\end{equation}
($r_m = m_\pi/m_K$). Thus  the two contributions $A_{IB}$ and
$A_{DE}^{electric}$ can interfere in the $E(z_i)$ amplitude,
contrary to the case of the amplitude $M(z_i)$
where only a direct emission contribution appears:
\begin{equation}
|A(K_{S,L} \rightarrow \pi^+\pi^-\gamma)|^2 =|A_{IB}|^2+2 \Re
\{ A_{IB}^*A_{DE}^{electric}\}+|A_{DE}^{electric}|^2+|A_{DE}^{magnetic}|^2.
\end{equation}

Finally the  magnetic and the electric direct emission  amplitudes can be
decomposed in a multipole expansion (see Refs. \cite{dibP,DMS}). Since higher
multipoles are suppressed by angular momentum barrier, in the following
we will consider only the lowest multipole component (the dipole one).
In this approximation the electric amplitude
is CP-conserving in the $K_S$ decay and CP-violating in the $K_L$ one,
while the magnetic amplitude is CP-conserving in the $K_L$ decay and
CP-violating in the $K_S$ one. For this reason, since $A_{IB}$ is enhanced by
the  pole at zero photon energy, the $K_S$ decay is completely
dominated by the electric transition, while
electric and magnetic contributions are of the same
order in the $K_L$ decay.

Similar to $K\to 2\pi$ and $K\to 3\pi$ cases it is convenient to
introduce the CP-violating parameter:
\begin{equation}
\eta_{+-\gamma}
=\frac{A( K_L\rightarrow \pi^+ \pi^-\gamma)_{IB+E1}}
{A( K_S\rightarrow \pi^+ \pi^-\gamma)_{IB+E1}},
\label{etagamma}
\end{equation}
where the subscript $E_1$ indicates that only the
lowest multipole component of the electric direct emission
amplitude has been considered. Using eq. (\ref{defib}) we can write:
\begin{equation}
\eta_{+-\gamma}= \eta_{+-} + \epsilon'_{\pi\pi\gamma}
\frac{A( K_S\rightarrow \pi^+ \pi^-\gamma)_{E1}}
{A( K_S\rightarrow \pi^+ \pi^-\gamma)_{IB+E1}}
\simeq
\eta_{+-} + \epsilon'_{\pi\pi\gamma}
\frac{A( K_S\rightarrow \pi^+ \pi^-\gamma)_{E1}}
{A( K_S\rightarrow \pi^+ \pi^-\gamma)_{IB}},
\label{epsprimepmg}
\end{equation}
where $\eta^{+-}$ is the usual $K\to 2\pi$  CP-violating parameter.
The term proportional to $\epsilon'_{\pi\pi\gamma}$
in eq. (\ref{epsprimepmg}) is a direct CP-violating
contribution, not related to the $K\to \pi\pi$ amplitude and consequently
not suppressed by the 1/22 factor of the $\Delta I=1/2$ rule.
However, although $\epsilon'_{\pi\pi\gamma}$ could be much larger than
$\epsilon'_{\pi\pi}$, the second term in eq. (\ref{epsprimepmg})
is suppressed by the factor $A_{E1}/A_{IB}\ll 1$.

The $\eta_{+-\gamma}$ parameter has already been measured at
Fermilab obtaining for the IB contribution \cite{Barker}
 \begin{equation}
\vert\eta_{+-\gamma (IB)}\vert=
\Big\vert \frac{A( K_L\rightarrow \pi^+\pi^-\gamma)_{IB}}
{A(K_S\rightarrow \pi^+ \pi^-\gamma)_{IB}}\Big\vert
 = (2.414 \pm 0.065 \pm 0.062 )\times 10^{-3},\label{etagammaf}
\end{equation}
\begin{equation}
\phi_{+-\gamma (IB)}=\arg (\eta_{+-\gamma (IB)})
 = (45.47 \pm 3.61 \pm 2.40)^\circ.
\label{phigamma}
\end{equation}
\DAFNE\ should improve these limits by studying
the time evolution of the decay.

Referring to \cite{NG} for an extensive analysis,
 here we show how to take advantage of the $\phi$-factories possibilities
to measure $\eta_{+-\gamma}$. Choosing as final states
 $f_1=\pi^\pm l^\mp \nu$, $f_2=\pi^+\pi^-\gamma$ and following the
notation of  Section 2, the time difference distribution,
integrated over final phase space, is given by\footnote{In the
following we neglect possible violations of CPT and of the $\Delta S =
\Delta Q$ rule.}:
\begin{eqnarray}
 I( \pi^{\pm} l^{\mp} \nu, \pi^+\pi^- \gamma; t < 0 ) =
\frac{\Gamma(K_L\to \pi^{\pm} l^{\mp} \nu)
\Gamma(K_S\rightarrow \pi^+\pi^-\gamma)}
{2\Gamma }
\Big\{ R_L e^{-\Gamma_L |t|}
+ e^{-\Gamma_S |t|} \nonumber \\
\pm \frac{2 \, e^{-\Gamma |t|}}{\Gamma(K_S\rightarrow \pi^+\pi^-\gamma)}
\Big[\Re \langle E\rangle_{int}
\cos({\Delta m |t|})
+\Im \langle E\rangle_{int}
\sin({\Delta m |t|})\Big]
\Big\},\label{sievolutionkppg}
\end{eqnarray}
where
$R_L=\Gamma(K_L\rightarrow \pi^+\pi^-\gamma)/
\Gamma(K_S\rightarrow \pi^+\pi^-\gamma)$
and
\begin{eqnarray}
\langle E \rangle_{int}\equiv & &
{m_K\over{(4\pi)^3}}\int\int dz_1 dz_2{(E_{IB}^S+E_1^S)^*
(E_{IB}^L +E_1^L)}\times \nonumber\\ & &
\times \big[ z_+ z_-(1-2z_3 - 2r_m^2)-r_m^2(z_+^2 +z_-^2)\big].
\label{interfkppg}
\end{eqnarray}
Neglecting the phase space dependence of $\eta_{+-\gamma}$ one should have
$\langle E \rangle_{int}
= \eta_{+- \gamma} \Gamma(K_S\rightarrow \pi^+\pi^-\gamma)$, and  therefore
the interference term of eq. (\ref{sievolutionkppg}) measures the CP-violating
$K_L \to \pi^+ \pi^- \gamma$ amplitude.
The expression for $t>0$ is obtained by interchanging $\Gamma_S\leftrightarrow
 \Gamma_L$ and changing
the sign of the $\sin({\Delta m |t|})$  term.
By fitting the experimental data with the theoretical expression of
eq. (\ref{sievolutionkppg}), one should be able to measure
the  interference term and then improve
the measurement of $\eta_{+-\gamma}$. Very useful to this purpose will be
the difference among the fluxes defined in eq. (\ref{sievolutionkppg})
 with positive and negative lepton charges, as discussed for $ K_S \to
3 \pi$ decays.

To conclude, we remark that not only the semileptonic tagging but
also the $K\to 2\pi$ one can be used to measure
$\eta_{+-\gamma}$ \cite{NG}.

\section{C-even background}
Some years ago it was pointed out that the radiative $\phi$ decay
 ($\phi \to \gamma f_0(980) \to \gamma (K^0 \bar K^0 )_{C=+}$)
could have a non-negligible branching ratio and therefore
could spoil the power of a $\phi$ factory in measuring $\rap$
and to single out the suppressed  $K_S$ decays \cite{cevena}.
 More recent determinations of the resonant contribution consider also the
$a_0(980)$, $f_0(980)$ interference effect and
 give a much lower value: the ratio
\begin{equation}
 r = \frac{\rm{Br}(\phi \to \gamma \, (K^0 \bar K^0)_{C+})}{\rm{Br}(\phi
\to (K^0 \bar K^0)_{C-})},
\end{equation}
is estimated to be in the range
$(3 \times 10^{-7})$--$(5 \times 10^{-9})$
 \cite{cevenb}.\par These predictions are strongly
model-dependent and larger values could perhaps  be obtained; however we can
trust that $r$ is certainly smaller than $10^{-4}$.
The  non-resonant contribution to $\phi \to \gamma \, K^0 \bar K^0$
 has been evaluated in the current algebra framework \cite{nello} to be
 of the order of $10^{-9}$,  comparable to the lower
 predictions of \cite{cevenb}.
   As we will show, the effects of the
C-even background on the \DAFNE~ measurements
are negligible also for unrealistically large values of $r$.\par
The C-even $K^0 \bar K^0$ state can be written as:
\begin{eqnarray}
  & &
{1\over \sqrt{2}}\big[ K^{0(q)} \bar K^{0(-q)} +  \bar K^{0(q)} K^{0(-q)}
\big]    \nonumber \\
& & \simeq
{1\over \sqrt{2}} \big[
K^{(q)}_S K^{(-q)}_S -K^{(q)}_L K^{(-q)}_L -
2 \Delta (K^{(q)}_S K^{(-q)}_L + K^{(q)}_L K^{(-q)}_S)\big],
\label{even}
\end{eqnarray}
where terms of order $\epsilon^2$  have been neglected but the
effect of possible CPT violation has been included.\par
The $K_S K_S$ component has a CP-conserving decay into
the $|2\pi, 2 \pi\rangle$
final states and could be a dangerous background in the $\rap$ measurement.
 However, the time difference distribution of
these events,
\begin{equation}
\overline {F(t)}=
         \frac{\Gamma^{+-}_{S}\Gamma^{00}_{S} }{2\Gamma_{S}}
        e^{-\Gamma_S |t|},
\end{equation}
is symmetric and cannot simulate the effect of $\epsilon'$.
Furthermore, it vanishes very rapidly for large values of $|t|$, and
therefore
it does not affect
 the determination of $\Re ( \rap )$. In effect,
 as shown in \cite{coco},   by means of a suitable cut the background
contribution can be eliminated in the
event sample used to determine $\Re ( \rap )$, also for very
 large values of $r$.

The background contribution  overlaps the signal just in the
interference zone,
$d \simeq d_{S}$, worsening the resolution on $\Im\left(\rap\right)$.
However the signal ($K_{L}K_{S} \rightarrow \pi^{0}\pi^{0},\pi^{+}\pi^{-}$)
and the background
($K_{S}K_{S} \rightarrow \pi^{0}\pi^{0},\pi^{+}\pi^{-}$)  have different
spatial behaviour, which is of help
in disentangling the signal contribution from the
background.
 The C-even background has been added to the signal in the
  fitting procedure of \cite{patera} and the accuracy achievable on
 $\Im\left(\rap\right)$ has been estimated again.
 The result is that for a realistic vertex
 resolution the worsening is around  5\%, even if $r$ would be as large
as $10^{-4}$.

The $K_L K_L$ component can affect the determination of the suppressed
$K_S$ branching ratios and the corresponding CP-violating asymmetries, as
we will discuss in the following.

If the $K_S$ semileptonic decays are
tagged as in  eq. (\ref{Nslpm}), the  number of events generated by
a single C-even $K^0 \bar K^0$ pair is:
\begin{equation}
\overline{N(l^\pm)}
 =  {\rm Br}(K_L \to l^\pm \pi^\mp \nu)
{\rm Br}(K_L \to x_L)
S_1 (1- e^{-10\Gamma_L/\Gamma_S}),
\end{equation}
then eq. (\ref{nslpmb}) is modified in:
\begin{equation}
N_S(l^\pm)_{exp} = 0.22 \times N_0 \left [ {\rm Br}(K_S \to l^\pm \pi^\mp \nu)
 +r\cdot  \frac{10 \Gamma_L}{\Gamma_S}{\rm Br}(K_L \to l^\pm \pi^\mp \nu)
\right],
\label{ncslpmb}
\end{equation}
 and the measured charge asymmetry becomes:
 \begin{equation}
(\delta_S)_{exp} =
\frac{N_S(l^+)_{exp} -N_S(l^-)_{exp}}{N_S(l^+)_{exp}+N_S(l^-)_{exp}} =
\frac{\delta_S + 10 r \delta_L}{1+ 10r}.
\end{equation}
As  can be seen the correction is absolutely negligible and cannot
simulate CPT violation, in fact $(\delta_S)_{exp} -\delta_L =(\delta_S -
\delta_L)/(1 +10 r)$.\par
The number of equal-sign dilepton events generated by
$(K^0 \bar K^0)_{C=+}$ is:
\begin{equation}
\overline {L^{++}} = {1 \over 2} \rm{Br}(K_L \to l^+ \pi^- \nu)
\rm{Br}(K_L \to l^+ \pi^- \nu)S_L \cdot S_L,
\end{equation}
and that of opposite sign is:
\begin{equation}
\overline {L^{+-}} =\overline {L^{-+}}=
 {1 \over 2} \rm{Br}(K_L \to l^+ \pi^- \nu)
\rm{Br}(K_L \to l^- \pi^+\bar{ \nu})S_L \cdot S_L.
\end{equation}
Therefore the experimentally measured T- and CPT-violating asymmetries are:
\begin{equation}
(A_{T})_{ exp} = \frac{\delta_S +\delta_L + 87 r (2 \delta_L)}{1 +  87 r}
\end{equation}
and
\begin{equation}
(A_{CPT})_{ exp} = \frac{A_{CPT}}{1 +  87 r}.
\end{equation}
Also in this case the effect of the C-even background is negligible
and the CPT prediction $A_{T} = 2 \delta_L$ is still valid. \par

 \begin{figure}[t]  
     \begin{center}
        \setlength{\unitlength}{1truecm}
        \begin{picture}(8.0,8.0)
           \put(-4.0,-6.0){\special{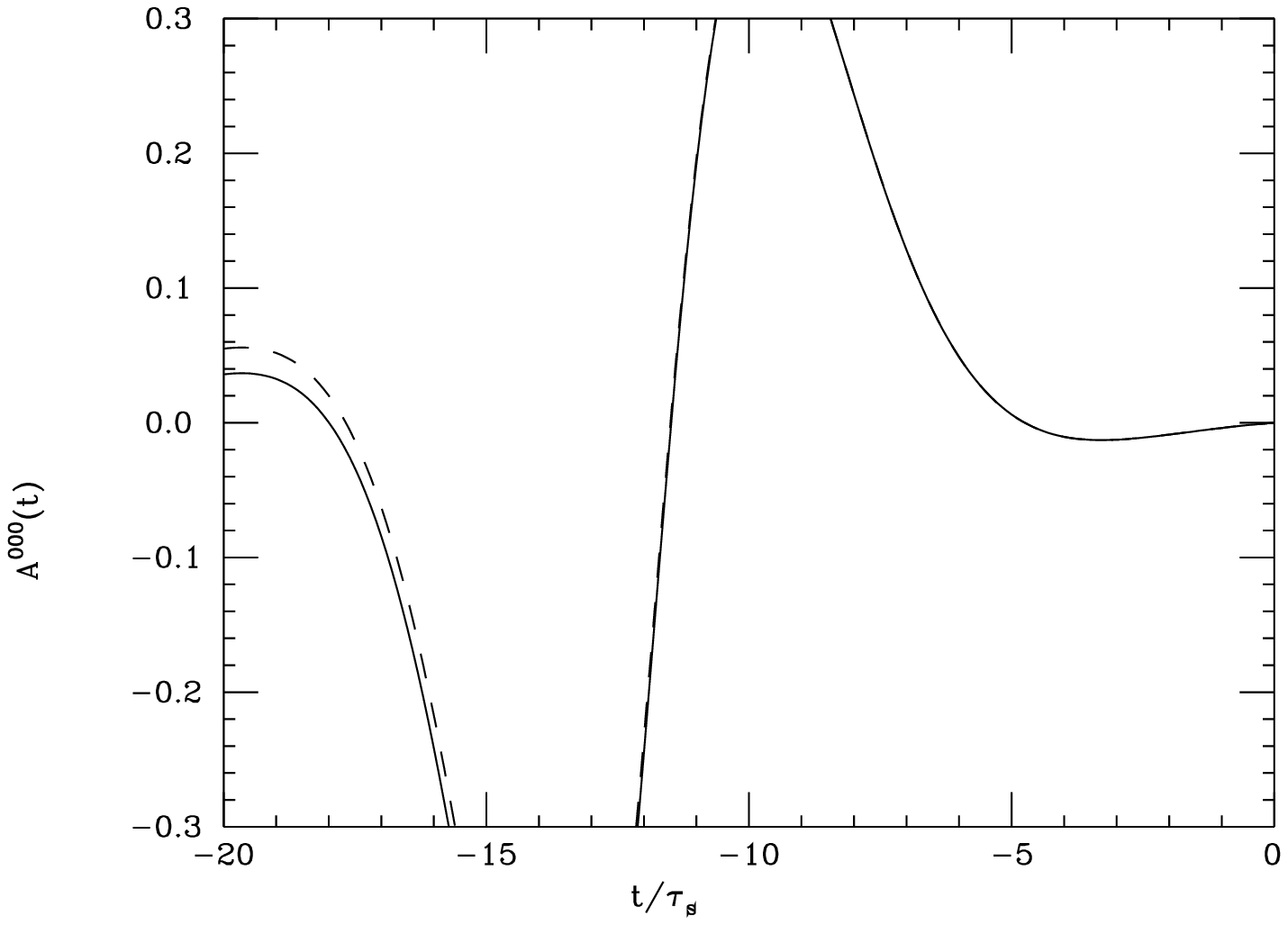}}
        \end{picture}
    \end{center}
    \caption{The effect of the C-even background on $A^{000}(t)$. The full
line corresponds to $r = 0$ and the dashed one to $ r =3 \times 10^{-7}$. }
    \protect\label{figceven}
\end{figure}

Finally we discuss the effect of C-even background in $ K_S \to 3\pi$
decays, where the largest influence is expected.\par
The inclusion of the background contribution in eq. (\ref{ksppp}) gives:
\begin{eqnarray}
& & (N_S(3\pi^0))_{exp} = N_0{\rm Br}(K_L \to 3 \pi^0)\times
\nonumber\\
& &\qquad \times\left[ 3.8 \times 10^{-4}|\eta^{000}|^2
 + 8.9 \times 10^{-10} - 1.8 \times 10^{-10}
\Im(\eta^{000})+ 3.8 \times 10^{-3} r \right].
\label{kspppc}
\end{eqnarray}
Therefore, for $r \geq 10^{-7}$  the background is comparable to the signal,
enforcing the conclusion that the direct tag of the $K_S$ is
not useful to determine Br($K_S \to (3 \pi)_{CP-}$).
On the contrary, if the  $K_S \to(\pi^+ \pi^- \pi^0)_{CP+}$
branching ratio is $ \sim 10^{-7}$, in agreement with ChPT predictions
 and preliminary CPLEAR results,
the background contribution to this decay
can be neglected and the corresponding branching ratio can be measured.

The modifications induced in interference measurements require a more
detailed study.
The analogue of the time difference distribution defined in eq. (\ref{intt})
for an initial $K_L\, K_L$ state is
\begin{equation}
\overline{I(a,b;t)} = {1\over 2 \Gamma_L}|A_L(a) A_L(b)|^2 e^{-\Gamma_L|t|}
(1 - e^{-2D/d_L +2 \Gamma_L|t|}),
\end{equation}
where the last factor accounts for having both $K_L$ decays inside
the detector.
Thus the C-even background, which is symmetric in $\pi^+ \pi^-$
momenta, modifies eqs. (\ref{a000}, \ref{apm0})  in the following way:
\begin{equation}
(A^{000}(t))_{exp} = \displaystyle \frac{ 2\Re \epsilon \left[
e^{+{\Delta \Gamma  \over 2}t} +  r {\Gamma \over \Gamma_L}
e^{+{\Delta \Gamma  \over 2}|t|}(1 - e^{-2D/d_L +2 \Gamma_L|t|})\right]
-2\Re\left(\eta^{000}e^{+i \Delta m t}\right) }{
 e^{+{\Delta \Gamma  \over 2}t} +|\eta^{000}|^2
 e^{-{\Delta \Gamma  \over 2}t} +
r {\Gamma \over \Gamma_L}
e^{+{\Delta \Gamma  \over 2}|t|}(1 - e^{-2D/d_L +2 \Gamma_L|t|})},
\label{ac000}
\end{equation}
and
\begin{equation}
(A^{+-0}(t))_{exp} = \displaystyle \frac{ 2\Re \epsilon \left[
e^{+{\Delta \Gamma  \over 2}t} +  r {\Gamma \over \Gamma_L}
e^{+{\Delta \Gamma  \over 2}|t|}(1 - e^{-2D/d_L +2 \Gamma_L|t|})\right]
-2\Re\left(\eta^{+-0}e^{+i \Delta m t}\right) }{
 e^{+{\Delta \Gamma  \over 2}t} +\frac{\Gamma_S^{+-0}}{\Gamma_L^{+-0}}
 e^{-{\Delta \Gamma  \over 2}t} +
r {\Gamma \over \Gamma_L}
e^{+{\Delta \Gamma  \over 2}|t|}(1 - e^{-2D/d_L +2 \Gamma_L|t|})}.
\label{cpm0}
\end{equation}
The contribution of the C-even background becomes important (especially in the
numerator) for large and negative values of the time difference,
where the number of events is absolutely negligible,
but, as can be seen from Fig. \ref{figceven},
does not affect the results for
 $ t \geq - 10 \tau_S $.

In the CP-conserving ratio $R^\pm(t)$ the C-even background
contributes in the denominator only and can be safely neglected.

\section{Quantum mechanics violations}

Research in the theory of quantum gravity has led to a proposal
for a  modification of quantum mechanics time evolution
\cite{QMV,ellis}, which might
transform a pure initial state into an incoherent mixture. This
effect becomes particularly interesting in the kaon system, where
quantum oscillations can be accurately measured. Moreover the
$\phi$-factory, where the initial $\bar{K}^0 K^0$
is an antisymmetric coherent state, is a very suitable facility to
test this idea.
We will not discuss the theoretical models of quantum mechanics
 violations
\cite{QMV,ellis} but, following the analysis of Ref. \cite{huet}, we
shall analyze some examples of observable effects at \DAFNE. \par

To describe the time evolution to incoherent states, one has to introduce
the formalism of the density matrix. In Ref. \cite{ellis} it has been
proposed to modify the quantum mechanics time evolution
equation in the following way:
\begin{equation}
i{{\rm d} \over {\rm d}t}\rho=H\rho-\rho H^\dagger + \widetilde{\cal H}\,
\rho, \label{qmvevol}
\end{equation}
where $\rho$ is the $2\times 2$ kaon density matrix, $H=M-i\Gamma/2$ is the
usual non-Hermitian kaon Hamiltonian (see eq. (18))
and $\widetilde{\cal H} \rho$ is
the quantum mechanics violating
term. For $\widetilde{\cal H}=0$ the eigenmatrices of eq. (\ref{qmvevol}) are
the usual matrices:
\begin{eqnarray}
&\rho_{LL}=|K_L\rangle\langle K_L|,\qquad &\rho_{SS}=|K_S\rangle\langle K_S|,
 \nonumber \\
&\rho_{SL}=|K_S\rangle\langle K_L|,\qquad &\rho_{LS}=|K_L\rangle \langle K_S|.
\label{eigenm}
\end{eqnarray}
Under reasonable assumptions (probability conservation,
not decreasing entropy and strangeness conservation)
$\widetilde{\cal H}$ can be expressed in terms of the three real parameters
$\alpha$, $\beta$ and $\gamma$ of Ref. \cite{ellis}. With this
parametrization the new eigenmatrices become:
\begin{eqnarray}
\widetilde{ \rho}_{LL}&=&\rho_{LL}+\left[ {\gamma \over \Delta\Gamma}
+ 4\beta{\Delta m \over \Delta \Gamma} \Im\left(
{\epsilon_L \over \Delta \lambda^*}\right)-{\beta^2\over |\Delta\lambda|^2}
 \right] |K_1\rangle\langle K_1| \nonumber \\ & &\qquad +
{\beta \over {\Delta \lambda}}|K_1\rangle\langle K_2|+
{\beta \over {\Delta \lambda}^*} |K_2\rangle\langle K_1|
 \nonumber \\
\widetilde{\rho}_{SS}&=&\rho_{SS}-\left[ {\gamma \over \Delta\Gamma}
+ 4\beta{\Delta m \over \Delta \Gamma} \Im\left(
{\epsilon_S \over \Delta \lambda^*}\right)-{\beta^2\over |\Delta\lambda|^2}
 \right] |K_2\rangle\langle K_2| \nonumber \\ & &\qquad -
{\beta \over {\Delta \lambda}^*}|K_1\rangle\langle K_2|+
{\beta \over {\Delta \lambda}} |K_2\rangle\langle K_1|
 \nonumber \\
\widetilde{\rho}_{SL}&=&\rho_{SL}-{\beta \over {\Delta \lambda}^*}
|K_1\rangle\langle K_1|+{\beta \over {\Delta \lambda}}|K_2\rangle\langle K_2|
-i{\alpha \over 2\Delta m}|K_2\rangle\langle K_1| \nonumber \\
\widetilde{\rho}_{LS}&=&\rho_{LS}-{\beta \over {\Delta \lambda}}
|K_1\rangle\langle K_1|+{\beta \over {\Delta \lambda}^*}|K_2\rangle\langle K_2|
-i{\alpha \over 2\Delta m}|K_1\rangle\langle K_2|,
\label{eigenmn}
\end{eqnarray}
where $|K_{1,2}\rangle$ are the usual CP eigenstates and
\begin{equation}
{\Delta \lambda}=\Delta m +i{\Delta\Gamma \over 2}=|\Delta \lambda
|e^{i(\pi/2 -\phi_{SW})}.
\end{equation}
The analysis of fixed target experiments has led
the authors of Ref. \cite{huet} to
put stringent bounds on the quantum mechanics
violating parameters $\beta$ and $\gamma$:
\begin{equation}
\begin{array}{rcl}
\beta&=& (0.32\pm 0.29)\times 10^{-18} {\rm  GeV},  \\
\gamma&=& (-0.2\pm 2.2)\times 10^{-21} {\rm  GeV}.
\end{array}
\label{qmvlimits}
\end{equation}
Quite similar results have been obtained in \cite{string}, where also a
bound for $\alpha$ ($\alpha \leq 4.8 \times 10^{-16}\ {\rm GeV}$) is derived.
It is interesting to note that, using the values in eq. (\ref{qmvlimits}),
the limits on $\alpha/m_K$ and $\beta/m_K$ turn  out to be of the order of
$m_K/M_{\rm Planck}$, which could be the natural suppression factor for
these parameters.

These limits have been obtained assuming  that there is no CPT
violation  in the decay amplitudes. In the more general case
the effects of $\beta$, $\gamma$ and those of
the conventional CPT-violating terms are mixed together.

This situation could be improved at \DAFNE. In effect,
 quantum mechanics predicts a vanishing
amplitude for the transition to the final state $|f(t_1),f(t_2)\rangle$,
with $t_1=t_2$, independently from possible CPT violations.
Therefore, as pointed out in
\cite{huet}, any measurement of equal time $f_1=f_2=f$ events can give
a bound  for pure quantum mechanics violations.
It should be stressed however that
the finite experimental resolution will partially wash out these effects
(see Fig. 2). Moreover, also the C-even background gives rise to equal time
 events.
 Thus only the time distribution, which is  different for
 C-even background and quantum mechanics violations,
could help in disentangling them.

\par
In Ref. \cite{huet} the consequences of quantum mechanics violation
to several \DAFNE\
observables have been analyzed. In particular the explicit formula
of the $|\pi^+\pi^-,\pi^0\pi^0\rangle$ time difference distribution
has been derived, discussing the quantum mechanics violating effects in the
measurement of $\rap$. For $t\gg \tau_S$ the time asymmetry of
eq. (\ref{aepsp}) becomes:
\begin{equation}
A(t) \stackrel{t\gg \tau_S}{\longrightarrow}
3 \Re (\rap) \left[ 1- {\gamma \over
\sqrt{2} |\Delta \lambda| |\epsilon|^2} -2
{\beta \over  |\Delta \lambda| |\epsilon|} \right]
-3  \Im (\rap) \left[2 {\beta \over  |\Delta \lambda| |\epsilon|} \right].
\end{equation}
In Fig. 7 we have plotted
$A(t)$ in the usual quantum mechanics case
and in the  quantum mechanics violating case (following the
analysis of  \cite{huet}), for $\Re(\rap ) = 5\times 10^{-4}$ and
$\Im (\rap )= -4\times 10^{-3}$, 0 and $+4\times 10^{-3}$ (close to
the predicted \DAFNE~  sensitivity).
The quantum mechanics violating parameters have been chosen to be:
$\beta= 0.71\times 10^{-18} {\rm  GeV} $ and $
\gamma=2.2\times 10^{-21} {\rm  GeV}$, in order to maximize their effects.
As one can see, for very small values of the time difference
$t$ the effects of quantum mechanics violations  are striking,
but probably beyond any realistic experimental resolution.
The determination of $\Im (\rap )$  should not be affected, but
an effect could be present in the asymptotic value of
the asymmetry where, contrary to the usual quantum mechanics case,
also $\Im(\rap)$ contributes.

\par
 \begin{figure}[t]
     \begin{center}
        \setlength{\unitlength}{1truecm}
        \begin{picture}(8.0,8.0)
        \put(-4.0,-6.0){\special{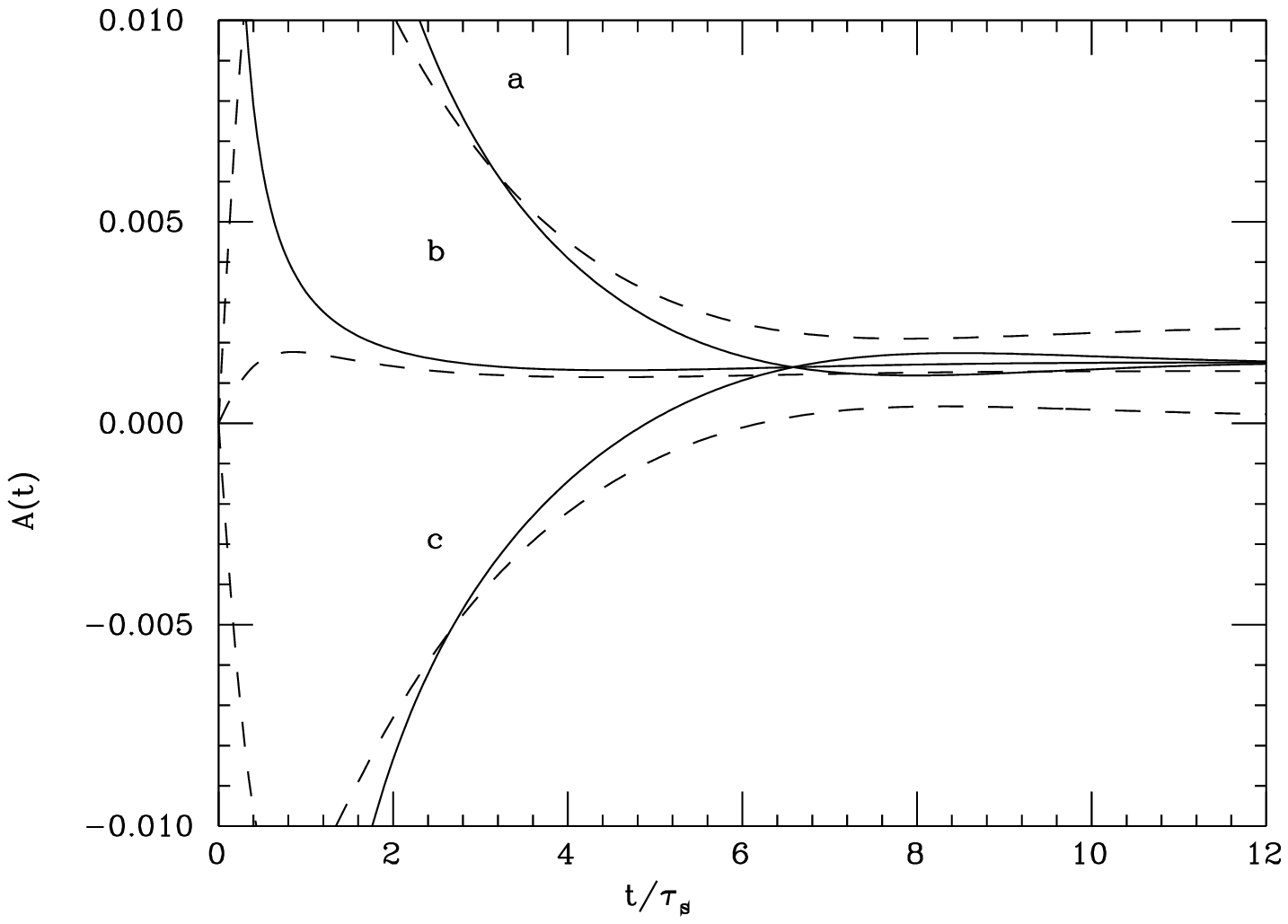}}
        \end{picture}
    \end{center}
    \caption{The time asymmetry $A(t)$ for $ \Re (\rap ) = 5\times 10^{-4}$.
The full lines correspond to quantum
mechanics predictions of eqs. (11) and (12),
with $ \Im (\rap ) =-4 \times 10^{-3}$ (a), $0$ (b) and
$4\times 10^{-3}$ (c).
The dashed lines correspond to the predictions of \protect{\cite{huet}}
with the upper bounds of quantum mechanics
violating parameters $\beta= 0.71\times 10^{-18} {\rm  GeV} $ and $
\gamma=2.2\times 10^{-21} {\rm  GeV}$,
for the previous values of $ \Im (\rap )$.}
    \protect\label{figqmv}
\end{figure}

Using the expressions of Ref. \cite{huet} and neglecting a
possible violation of the $\Delta S=\Delta Q$ rule, we have calculated
also the quantum mechanics violating effects for eq. (\ref{Nslpm}).
The measured charge asymmetry for $K_S$ semileptonic decays would become:
\begin{equation}
(\delta_S)_{exp}
  \equiv  \frac{(N_S(l^+))_{exp}-(N_S(l^-))_{exp}}
{(N_S(l^+))_{exp}+(N_S(l^-))_{exp}}=
\frac{\delta_S
   -2\Re\left({\beta\over {\Delta \lambda}}\right)
   +4\sqrt{2}{\beta\over |{\Delta \lambda}|}
   +20{\gamma\over \Delta\Gamma}\delta_L }{1+20{\gamma \over \Delta\Gamma}},
\label{QMVds}
\end{equation}
where $\delta_{S,L}$ are the usual asymmetries, as defined in
eq. (\ref{deltasl}).  The effect of  the $\gamma$ term simulates
a C-even background and interestingly, with the
limit of eq. (\ref{qmvlimits}), this correction turns out to be
of the order of the one estimated in the previous section
(for $r\sim 10^{-7}$). The  $\beta$ term
simulates a CPT violation, however, using
the bound in eq. (\ref{qmvlimits}), this effect turns out to be
smaller than the \DAFNE~ sensitivity.
\par
Concluding, we can say that quantum mechanics violating effects can be
neglected in integrated asymmetries at \DAFNE. Nevertheless,
if $\alpha$, $\beta$ and $\gamma$ were suppressed only linearly by
$m_K/M_{\rm Planck}$, some
effects in time-dependent distributions could be observable.
An estimate of \DAFNE~ sensitivity on these effects
would require an accurate simulation of the experimental apparatus,
which is beyond the purpose of this work.

\section{Conclusions}
\par
{}From the previous analysis it is clear that a $\phi$ factory is very
suitable for an accurate study of the origin of CP violation in $K_S$ and
$K_L$ decays and to test the Standard Model predictions.\par
The real part of the ratio \rapp which is a clear signal of direct CP
 violation,
can be measured with high precision, about $10^{-4}$. A non-vanishing value
of the imaginary part of \rapp which would imply CPT violation, can be
detected up to some units in $10^{-3}$. Even if the fixed target experiments
will reach a similar sensitivity, the KLOE apparatus has a completely
different systematics and such experimental result will be very important.

The presence of a pure $K_S$ beam will allow, for the first time, the direct
detection of CP violation in $K_S$ decays. Moreover many interesting tests
of T and CPT symmetries (in addition to $\Im (\rap)$) can be performed.\par
The combined analysis of \DAFNE, CPLEAR, E731 and NA31 experiments will
allow to disentangle the CPT-violating contributions in decay amplitudes
from those  in mass matrix and also the $ \Delta S = - \Delta Q$ transitions
 can be singled out. All the real parts of the
parameters could be bounded up to $ \simeq 5 \times 10^{-4}$
while a lower sensitivity ($\sim 10^{-3}$--$10^{-2}$)
is expected for the imaginary parts.\par
More doubtful is the situation for $K_S \to 3 \pi$ decays. We think
that the CP-conserving decays will certainly be measured and some
information  on the rescattering phases could be obtained. For the
CP-violating ones we observe, without doing a complete statistical analysis,
that the shape of the interference effect of Fig. 4 is very
characteristic and could easily be detected over a flat background.
The analysis of the possibility to measure $\eta_{+-\gamma}$ leads to
similar conclusions.

Also the recent suggestions on possible quantum mechanics
violations might be tested at \DAFNE.

\vskip 0.5 cm
\noindent
{\bf Acknowledgements}
\par
\noindent
We wish to thank V. Patera and N. Paver for very useful
suggestions and comments. We acknowledge also the members of
the DA$\Phi$NE working groups for interesting discussions.

\end{document}